%% file: Main.tex
\documentclass[preprint,pteplogo]{ptephy_v2}

\usepackage{hyperref}
\usepackage[utf8]{inputenc}
\usepackage[T1]{fontenc}
\usepackage{lineno}
\usepackage{multirow}

\usepackage{graphicx}
\usepackage{tabularx}
\usepackage{siunitx}
\usepackage{threeparttable}
\newtheorem{theorem}{Theorem}
\newtheorem{condition}{Condition}
\input{MyCommands}

\begin{document}

\title{Measurement of reconstructed final-state kinematics\\in charged-current interactions on water\\using the J-PARC $\nu_\mu$ beam and NINJA emulsion detector}


\input{SubFiles/Authors}





\input{SubFiles/0_Abstract}

\subjectindex{C32, D02}
\maketitle
\input{SubFiles/1_Introduction}
\input{SubFiles/2_Beam_Detectors}
\input{SubFiles/3_Simulation}
\input{SubFiles/4_Analysis}
\input{SubFiles/5_Systematics}
\input{SubFiles/6_Results}
\input{SubFiles/7_Conclusions}

\input{SubFiles/Acknowledgment}

\bibliographystyle{ptephy}
\bibliography{SubFiles/References}
\end{document}

%% file: MyCommands.tex
\usepackage[compat=1.1.0]{tikz-feynhand}

\newcommand{\memo}[1]{\marginpar{\footnotesize{#1}}}
\renewcommand{\memo}[1]{}	

\newcommand{\Fig}[4]{
	\begin{figure}[htbp]
		\begin{center}
            \includegraphics[keepaspectratio, width=#2]{Fig/#1}
			\caption{#3}
			\label{#4}
		\end{center}
	\end{figure}
}


%% file: SubFiles/Authors.tex

\newcommand{\AFFtoho}{\affil{Department of Physics, Toho University, Funabashi 274-8510, Japan}}

\newcommand{\AFFkyoto}{\affil{Department of Physics, Kyoto University, Kyoto 606-8502, Japan}}

\newcommand{\AFFkobe}{\affil{Graduate School of Human Development and Environment, Kobe University, Kobe 657-8501, Japan}}

\newcommand{\AFFtitech}{\affil{Department of Physics, Institute of Science Tokyo, Tokyo 152-8551, Japan}}

\newcommand{\AFFnagoya}{\affil{Department of Physics, Nagoya University, Nagoya 464-8602, Japan}}

\newcommand{\AFFnihon}{\affil{College of Industrial Technology, Nihon University, Narashino 275-8576, Japan}}

\newcommand{\AFFkamioka}{\affil{Kamioka Observatory, Institute for Cosmic Ray Research, The University of Tokyo, Kamioka 506-1205, Japan}}

\newcommand{\AFFtohoku}{\affil{Department of Physics, Tohoku University, Sendai 980-8578, Japan}}

\newcommand{\AFFyokohama}{\affil{Faculty of Engineering, Yokohama National University, Yokohama 240-8501, Japan}}

\newcommand{\AFFkanagawa}{\affil{Faculty of Engineering, Kanagawa University, Yokohama 221-8686, Japan}}

\newcommand{\AFFtokyo}{\affil{Department of Physics, The University of Tokyo, Tokyo 113-0033, Japan}}

\newcommand{\AFFriken}{\affil{Nishina Center for Accelerator-Based Science, RIKEN, Wako 351-0198, Japan}}

\newcommand{\AFFtokyodenki}{\affil{School of Science and Engineering, Tokyo Denki University, Ishizaka, Saitama 350-0394, Japan}}

\newcommand{\AFFrbi}{\affil{Center of Excellence for Advanced Materials and Sensing Devices, Ru{\dj}er Bo{\v{s}}kovi{\'c} Institute, 10000 Zagreb, Croatia}}

\newcommand{\AFFkcl}{\affil{King’s College London, London, WC2R 2LS, United Kingdom}}

\newcommand{\AFFosaka}{\affil{Research Center for Nuclear Physics, Osaka University, Osaka 567-0047, Japan}}

\newcommand{\AFFipmu}{\affil{Kavli Institute for the Physics and Mathematics of the Universe, The University of Tokyo, Tokyo 277-8583, Japan}}


\newcommand{\industrymark}{\textsuperscript{\dag}}

\author[1,*]{{A.~Kasumi}}
\email{kasumi@flab.phys.nagoya-u.ac.jp}
\AFFnagoya

\author[2,*]{{S.~Han}}
\email{han.seungho.3s@kyoto-u.ac.jp}
\AFFkyoto

\author[3]{E.~Abad Díaz}
\AFFrbi

\author[4]{S.~Aoki}
\AFFkobe

\author[3]{{D.~Barčot}}

\author[5]{{C.~Bronner}}
\AFFyokohama

\author[1,2]{{T.~Fukuda}}
\author[1]{{Y.~Furuta}}
\author[3]{{M.~Ghosh}}
\author[1]{{T.~Hayakawa}}
\author[1]{{Y.~Hayasaka}\thanks{Currently in industry.}}

\author[6]{{Y.~Hayato}}
\AFFkamioka

\author[3]{{L.~Halić}}
\author[5]{D.~Hirata}
\author[1]{H.~Hotta}
\author[3]{{M.~Jakkapu}}

\author[7]{C.~Jesús-Valls\thanks{Present address: CERN, Geneva, Switzerland}}
\AFFipmu

\author[8]{{T.~Katori}}
\AFFkcl

\author[1]{{H.~Kawahara}\industrymark}
\author[1]{{T.~Kawahara}}
\author[2]{{T.~Kikawa}}
\author[3]{{B.~Kilček}}
\author[1]{{H.~Kobayashi}}
\author[1]{{R.~Komatani}}
\author[1]{{M.~Komatsu}}
\author[3]{{B.~Kovač}}
\author[1]{{T.~Matsuo}}

\author[9]{{S.~Mikado}}
\AFFnihon

\author[5]{{A.~Minamino}}
\author[10]{K.~Mizuno\industrymark}
\AFFtoho
\author[1]{{Y.~Morimoto}}
\author[1]{K.~Morishima}
\author[1]{Y.~Nakamura}
\author[1]{T.~Nakano}

\author[2]{{T.~Nakaya}}

\author[1]{{T.~Nishikiori}}

\author[1]{H.~Oaira}
\author[2]{T.~Odagawa\industrymark}

\author[10]{S.~Ogawa}

\author[11]{H.~Oshima}
\AFFtokyodenki

\author[2]{{N.~Otani}}
\author[5]{G.~Pintaudi\industrymark}
\author[1]{H.~Rokujo}
\author[1]{O.~Sato}

\author[12]{H.~Shibuya}
\AFFkanagawa

\author[1]{{S.~Shimizu}}

\author[1]{K.~Sugimura}
\author[1]{Y.~Suzuki\industrymark}

\author[1]{{S.~Takeshita}}
\author[2]{K.~Yasutome\thanks{Present address: RIKEN SPring-8 Center, Hyogo, Japan}}

\author[1]{S.~Yamamoto}
\author[13]{M.~Yoshimoto}
\AFFriken

\collaborator{NINJA Collaboration}


%% file: SubFiles/0_Abstract.tex
\begin{abstract}

We present a study of charged-particle multiplicities and the kinematic distributions of muons, protons, and charged pions in a flux-integrated sample of charged-current inclusive $\nu_\mu$ interactions on water recorded with the NINJA nuclear emulsion detector exposed to the J-PARC $\nu_\mu$-focused beam. The data correspond to an exposure of $4.63\times10^{20}$ protons on target, with a neutrino energy spectrum peaked at 0.7~GeV. In this analysis, 82 events are selected from a fiducial water volume with a mass of 3.9~kg, corresponding to 5\% of the total water target mass exposed during the run. The reconstructed distributions are compared with Monte Carlo predictions based on the interaction model used in the T2K experiment. The observed distributions are generally consistent with the predictions within the estimated uncertainties, although the proton angular distribution exhibits some tension with the prediction. The results are currently limited by statistical uncertainties. This analysis provides the first study of proton kinematics with sensitivity to proton momenta as low as 200\ MeV/$c$ with a $\nu_\mu$-focused beam on a water target using nuclear emulsion data and establishes the analysis framework for future measurements with substantially larger NINJA data samples.

\end{abstract}

%% file: SubFiles/1_Introduction.tex
\section{Introduction}

Modeling neutrino--nucleus interactions plays a central role in long-baseline neutrino oscillation experiments, where the incident neutrino energy is reconstructed from the particles observed in the detector. Nuclear effects, including nucleon motion, nucleon--nucleon correlations, and final-state interactions, affect the observable final state and introduce significant uncertainties in the relationship between the true and reconstructed neutrino energies \cite{lcpv_review, hk_sensitivity, missing_energy}. Accurate modeling of these effects is therefore essential for precision measurements of leptonic CP violation and the determination of the neutrino mass ordering \cite{nustec}.

Measurements of outgoing hadron kinematics and their correlations with lepton kinematics provide a powerful probe of nuclear effects. Compared with electron--nucleus scattering, however, such measurements in neutrino interactions are limited by small interaction cross sections and broad neutrino energy spectra, leaving the axial nuclear response relatively poorly constrained. Furthermore, the reconstruction of low-momentum hadrons, particularly protons, remains challenging for conventional tracking detectors.

Several experiments have recently reported measurements of final-state hadrons in neutrino interactions, including MicroBooNE on argon \cite{ub_2p, ub_first_2diff_tki, ub_gki, ub_inclusive_proton_prd, ub_simultaneous_proton_prl} and MINERvA \cite{mv_tki_prl_2018, mv_tki_2020}, NOvA \cite{nova_evis_hadron}, and T2K \cite{t2k_tki, t2k_tki_1pip} on hydrocarbon targets. In contrast, measurements of final-state hadron kinematics on water targets remain scarce, largely due to the challenge of reconstructing low-momentum hadrons, despite the importance of water targets for current and future neutrino oscillation experiments such as T2K \cite{t2k_2011} and Hyper-Kamiokande \cite{hk_sensitivity}.

The NINJA experiment addresses this challenge using nuclear emulsion detectors, which can detect protons with momenta down to about $200~\mathrm{MeV}/c$ and feature a modular design that enables the use of various target materials, including water. Combined with the intense J-PARC neutrino beam, NINJA is well suited for detailed studies of low-momentum hadrons in neutrino--water interactions. Previous pilot runs have demonstrated this capability, including measurements with a $1.2~\mathrm{GeV}$ $\nu_\mu$-focused beam on a 65 kg stainless steel (SUS304/AISI 304) target \cite{oshima_ptep, oshima_prd} and a $1.0~\mathrm{GeV}$ $\bar{\nu}_\mu$-focused beam on a 3 kg water target \cite{hiramoto_prd}, both using the INGRID detector \cite{ingrid} as a downstream muon detector.

\begin{table}[hb]
\centering
\caption{Summary of NINJA measurements, including beam mode, peak neutrino energy ($E_\text{peak}$), integrated POT, target material and fiducial mass, and the associated muon detector.}
\label{tab:ninja_papers}
\resizebox{\textwidth}{!}{%
\begin{tabular}{lcccccc}
\hline
\hline
 & Beam mode & $E_\text{peak}$ [GeV] & POT [$10^{20}$] & Target (mass in kg)  & $\mu$ detector\\
\hline
Refs.~\cite{oshima_ptep, oshima_prd} & $\nu_\mu$ & 1.2 & 0.4 & SUS304 (65)  & INGRID \cite{ingrid}\\
Ref.~\cite{hiramoto_prd} & $\bar{\nu}_\mu$ & 1.0 & 7.1 & Water (3.0) & INGRID \cite{ingrid}\\
This work & $\nu_\mu$ & 0.7 & 4.6 & Water (3.9)  & BabyMIND \cite{babymind}\\
\hline
\hline
\end{tabular}}
\end{table}

In this work, we present a reconstructed-level\footnote{Throughout this paper, ``reconstructed-level'' refers to a direct comparison between reconstructed data and reconstructed predictions, without unfolding the data to the true interaction level.} study of a selected sample of flux-integrated charged-current (CC) inclusive $\nu_\mu$ interactions on a $3.9~\mathrm{kg}$ water target using a $\nu_\mu$-focused beam with a peak energy of $0.7~\mathrm{GeV}$ and an exposure of $4.63 \times 10^{20}$ protons on target (POT). The analyzed sample corresponds to approximately 5\% of the first NINJA physics run data, accumulated from November 2019 to February 2020 with a total water target mass of 74 kg. This subset was selected for the first complete analysis, while quality assessment and processing of the remaining data are ongoing.

A comparison with previous NINJA pilot run measurements is summarized in Table~\ref{tab:ninja_papers}. In NINJA analyses, muon candidates are identified by matching emulsion tracks to those reconstructed in the downstream electronic muon detector. The physics run employs a new downstream muon tracking system consisting of the BabyMIND detector \cite{babymind} and two intermediate detectors, the Large Emulsion Shifter and the Scintillation Tracker \cite{odagawa_tracker} (see Secs.~\ref{sec:les_st} and \ref{sec:bm} for details). Compared with the pilot run configuration, this system enhances muon track matching performance and extends the muon angular acceptance to include the previously inaccessible region from $30^\circ$ to $50^\circ$ with respect to the beam. Improvements in microscope scanning and track connection algorithms within the emulsion detector \cite{suzuki_large_angle} further extend the hadron angular acceptance by incorporating tracks with angles of approximately $50^\circ$--$75^\circ$ and $105^\circ$--$130^\circ$ with respect to the beam.

Within this framework, reconstructed multiplicities and kinematic distributions of final-state muons, protons, and charged pions are compared with Monte Carlo (MC) predictions based on the interaction model configuration used in the T2K experiment \cite{t2k_epjc_2023}. The primary objective of this analysis is to establish the track reconstruction and analysis framework for the NINJA physics run using a limited subset of the collected data. 

Also, this work provides the first reconstructed-level study of proton kinematics down to a momentum threshold of about $200~\mathrm{MeV}/c$ in $\nu_\mu$ interactions on water using a neutrino beam with a spectrum peaked at $0.7~\mathrm{GeV}$, close to the energy range most relevant for oscillation measurements in T2K and Hyper-Kamiokande. At these energies, final-state particles are more frequently produced at low momenta and large angles, making the improved tracking capabilities developed for this analysis particularly important. 

%% file: SubFiles/2_Beam_Detectors.tex
\section{Neutrino beam and detector configuration}

\Fig{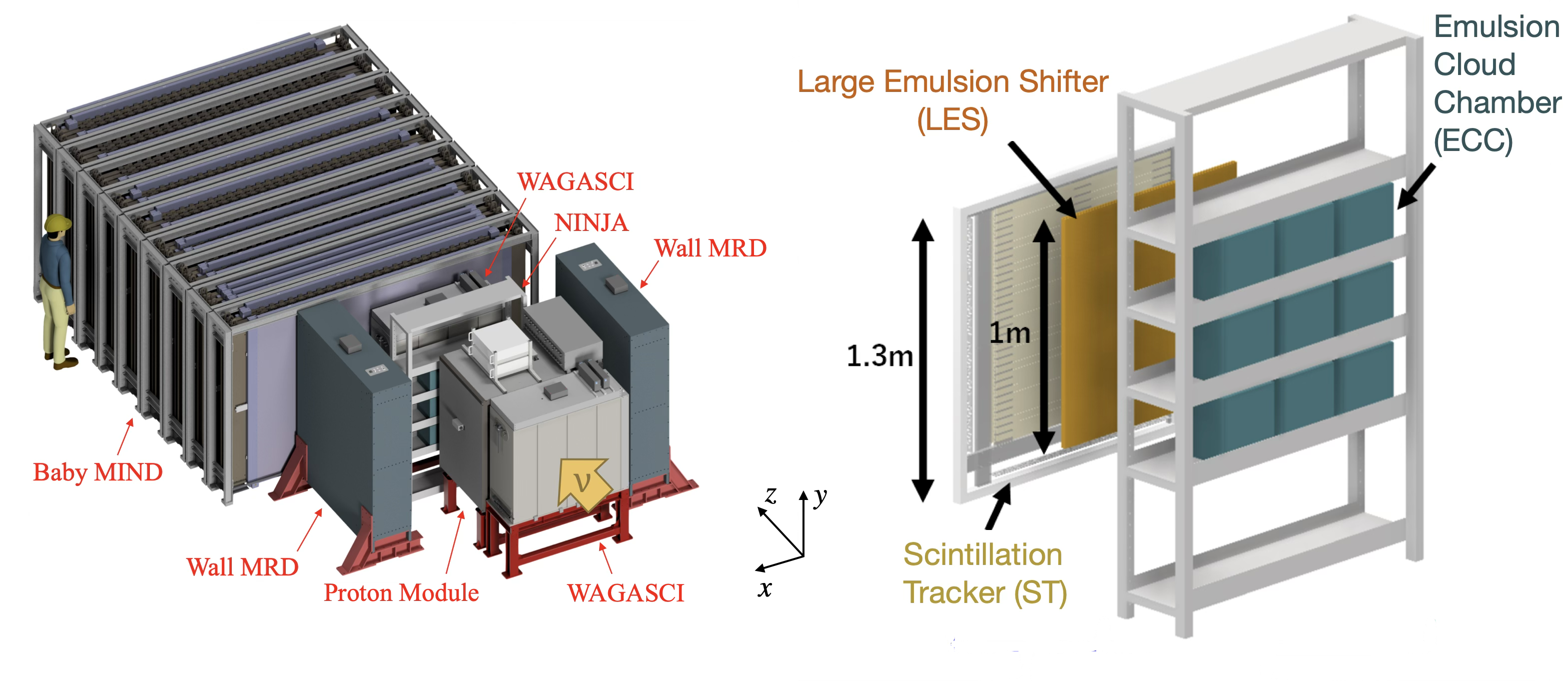}{\textwidth}{Placement of the NINJA detectors within the WAGASCI--BabyMIND setup (left), and a schematic view of the NINJA detector system used in the first NINJA physics run (November 2019--February 2020), including the \(3\times3\) arrangement of ECC modules (right).}{fig:detectors}

The NINJA detectors were installed on the B2 floor of the Neutrino Monitor building at J-PARC, between the two WAGASCI detector modules~\cite{wgbm_2025} of the T2K experiment. They were located 280~m downstream of the graphite target of the J-PARC neutrino beamline at an off-axis angle of $1.5^\circ$.

Figure \ref{fig:detectors} shows the schematic of the detector components and their placement within the WAGASCI-BabyMIND detectors \cite{wgbm_2025}. In this analysis, BabyMIND (BM) was used to identify particle tracks coincident with the beam spill, mostly muons from $\nu_\mu$ and $\bar{\nu}_\mu$ CC interactions, while the Proton Module (PM) \cite{proton_module} was used to identify muons originating from upstream.

\subsection{J-PARC muon neutrino beam}

The data set used in this analysis corresponds to the beam exposure accumulated from November 8, 2019 to February 12, 2020, totaling $4.63\times10^{20}$ POT. During this period, the J-PARC Main Ring reached a maximum beam power of 515~kW and delivered up to approximately $2.66\times10^{14}$ protons per spill. Each spill consisted of eight bunches with a width of 58~ns, separated by 580~ns, with a repetition cycle of 2.48~s. 

A 30~GeV proton beam impinges on a graphite target, producing secondary hadrons that are focused by a system of magnetic horns. By reversing the horn polarity, either a $\nu_\mu$ or $\bar{\nu}_\mu$-focused beam can be produced. For this data set, the horn current was set to 250~kA in neutrino mode. Further details of the beamline configuration can be found in Ref.~\cite{t2k_epjc_2023}.

Figure \ref{fig:nu_flux} shows the neutrino flux components simulated at the detector location. The $\nu_\mu$ purity of the beam is predicted to be approximately 92\%, with contamination of $\bar{\nu}_\mu$ ($6\%$), $\nu_e$ ($1\%$), and $\bar{\nu}_e$ (below 1\%). The neutrino flux is expected to peak at 0.70~GeV, with a mean energy of 1.04~GeV.

\Fig{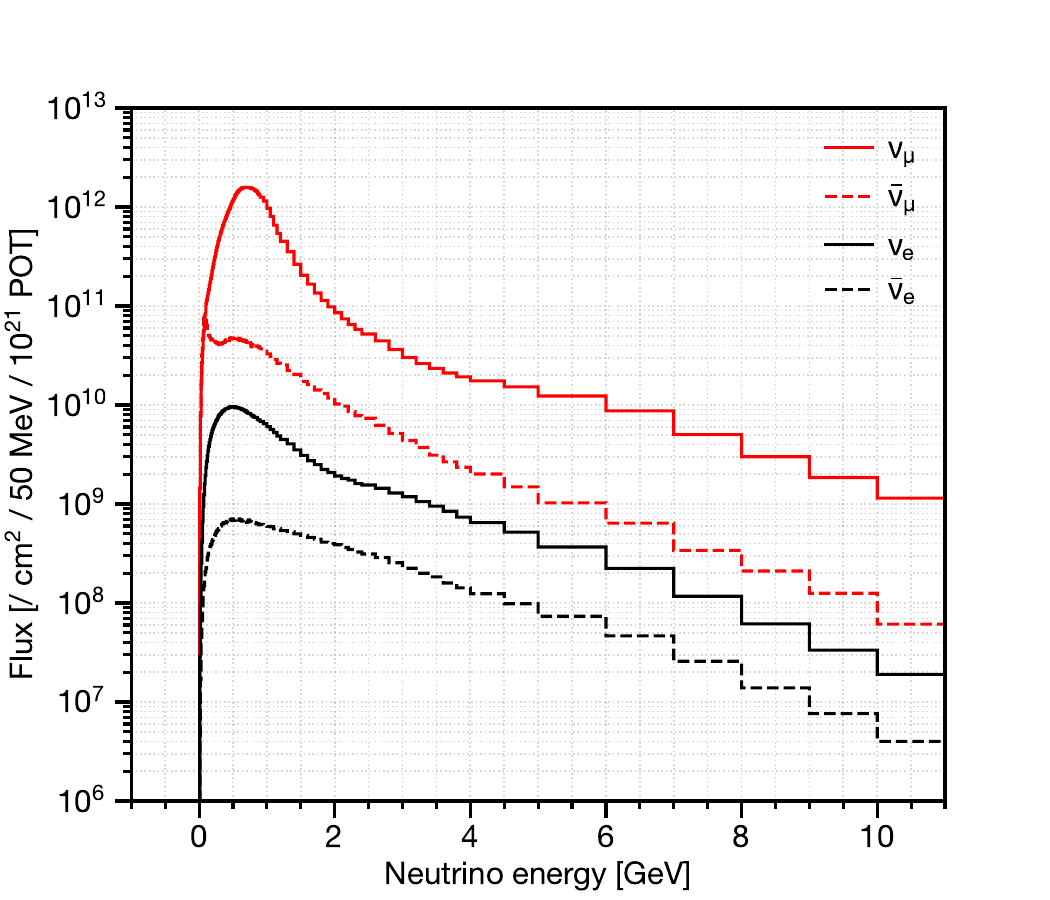}{0.7\textwidth}{\textcolor{black}{Expected neutrino flux spectra at the detector location.}}{fig:nu_flux}

\subsection{Neutrino interaction vertex detector: Emulsion Cloud Chamber (ECC)}

\Fig{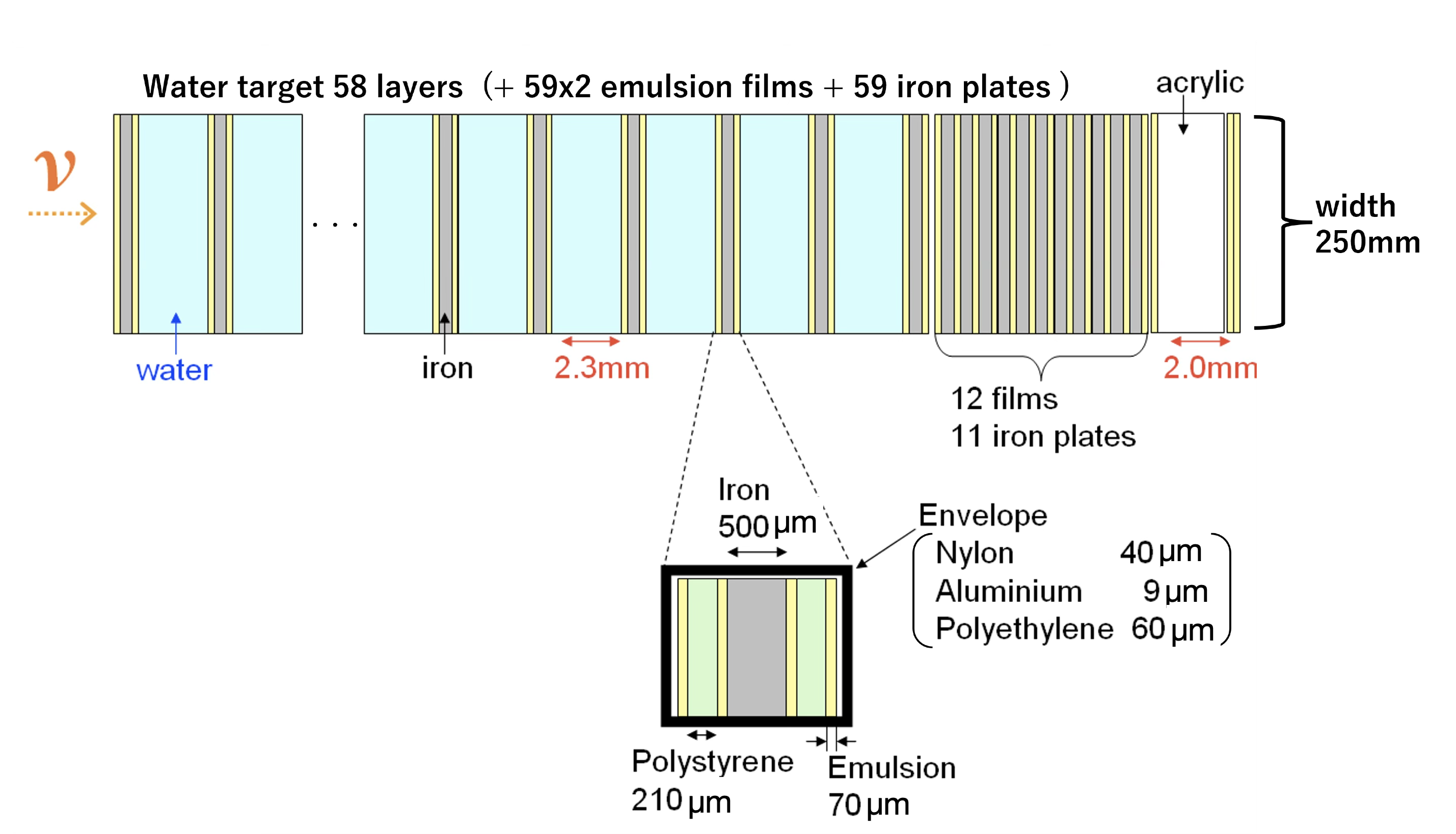}{0.9\textwidth}{Schematic illustration of the Emulsion Cloud Chamber (ECC), including its principal dimensions. Not to scale.}{fig:ecc}

Nuclear emulsion is a tracking medium with exceptional spatial resolution, in which silver bromide crystals are dispersed within a gelatin matrix. Emulsion films are produced by coating both sides of a polystyrene base with sensitive emulsion layers. An Emulsion Cloud Chamber (ECC), constructed by alternating such films with target material, provides three-dimensional tracking capability with sub-µm spatial resolution. Figure~\ref{fig:ecc} shows the ECC structure used in the first NINJA physics run.

Each tracking unit comprised a 500~\textmu m-thick stainless steel (SUS316L/AISI 316L) plate placed between two nuclear emulsion films (area: $25 \times 25~\mathrm{cm}^2$, thickness: 350~\textmu m; hereafter referred to as ``film''), and was vacuum-packed in a light-tight bag (``envelope'') consisting of nylon, aluminum, and polyethylene layers. A total of 59 tracking units were stacked in an acrylic box of $30 \times 30 \times 30~\mathrm{cm}^3$. Adjacent units were separated by 2.3~mm water gaps, resulting in 58 water layers per ECC with a total water mass of 8.2~kg (``water target section''). An additional 12 emulsion films and 11 steel plates were installed without gaps to form the downstream tracking section, which provides additional material for momentum reconstruction through multiple Coulomb scattering (MCS) in the steel plates.

Nine ECCs were mounted on a rack in a $3 \times 3$ configuration. In this analysis, only the central ECC was used as a first demonstration of the detector performance and analysis methodology; studies of the remaining ECCs are ongoing. For each emulsion film, the fiducial region used for neutrino vertex reconstruction was restricted to the central $17 \times 17~\mathrm{cm}^2$ area, leaving a 4~cm margin on each side to ensure sufficient track length for hadrons emitted at large angles. The fiducial volume was defined exclusively within the water target section.

\subsection{Muon detector: BabyMIND (BM)}
\label{sec:bm}

BabyMIND (BM) is a segmented detector consisting of alternating magnetized iron plates and plastic scintillator planes, forming a sampling spectrometer. The scintillator bars are read out via wavelength-shifting (WLS) fibers coupled to silicon photomultipliers (SiPMs), providing beam-bunch-level timing and $\mathcal{O}(1)$~cm spatial resolution for track reconstruction, enabling matching to upstream detectors. BM serves as the downstream muon detector in this analysis, providing muon identification and timing information. Unlike the WAGASCI--BabyMIND analysis~\cite{wgbm_2025}, neither the charge-sign information nor the range-based momentum measurement was used in this analysis. Instead, the measured track range was used only for muon identification, while the muon momentum was reconstructed using MCS in the ECC, as described in Sec.~\ref{sec:mcs_mom}. The detector design and performance are described in Refs.~\cite{babymind, wgbm_2025}.

\subsection{Intermediate detectors: Large Emulsion Shifter (LES) and Scintillation Tracker (ST)}
\label{sec:les_st}

The ECC records all charged particle tracks accumulated from film production to development, without intrinsic timing information. Over the roughly three-month exposure period, each emulsion film accumulates a track density of $\mathcal{O}(10^4)/\mathrm{cm}^{2}$. To associate these tracks with a specific beam bunch, all tracks reconstructed in the ECC were matched to corresponding tracks in the downstream muon detector, BM. This requires precise timing and position measurements to identify the correct track among the large number of unrelated tracks recorded in the emulsion films. Two intermediate detectors, the Large Emulsion Shifter (LES) and the Scintillation Tracker (ST), provide complementary timing and spatial information.

\Fig{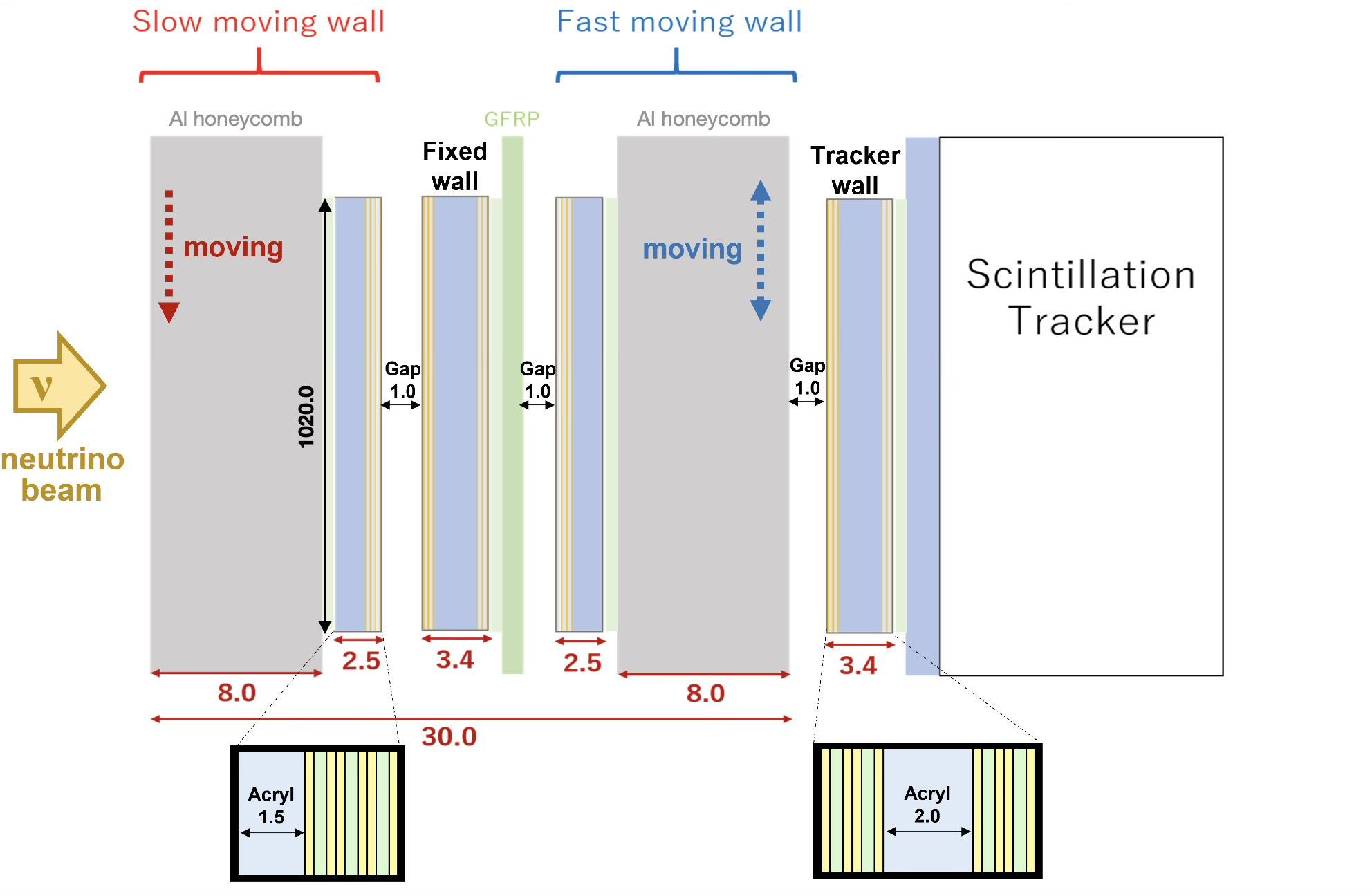}{\textwidth}{Schematic top-view illustration of the Large Emulsion Shifter (LES), showing its design, principal dimensions, and placement relative to the Scintillation Tracker (ST). Not to scale. All dimensions are given in mm.}{fig:les_st} 

The Large Emulsion Shifter (LES) provides coarse timing information. It consists of emulsion films mounted on three mechanically independent stages: an upstream slow-moving wall (displaced in 2~mm steps every 4 days), a fixed wall, and a downstream fast-moving wall (displaced in 2~mm steps every 4 hours). Each film has an active area of $1 \times 1~\mathrm{m}^2$. A traversing charged particle leaves tracks in the three walls with relative displacements determined by the stage positions, thereby encoding timing with a 4-hour resolution. This timing information restricts the search for the emulsion track associated with a given downstream muon candidate to a single 4-hour time window, reducing the density of candidate tracks and the probability of accidental track associations by a factor of $\mathcal{O}(10^{-3})$. The geometry of the films and stages, as well as their placement relative to the ECC, are illustrated in Fig.~\ref{fig:les_st}.

Downstream of the LES, the Scintillation Tracker (ST) provides precise timing at the \textcolor{black}{beam spill} level. The ST has two orthogonal planes of scintillator bars, oriented horizontally and vertically. Each plane consists of four layers of plastic scintillator bars (248 channels in total), each with a thickness of 3~mm, a width of 24~mm, and a length of 1~m. The bars are arranged with partial overlap to eliminate inactive regions and achieve a spatial resolution finer than the bar width alone. Each bar is read
out through a WLS fiber coupled to a SiPM. A
hit is defined as a signal above a threshold of 2.5 photoelectrons. For tracks with incident angles below $45^\circ$, the spatial resolution is better than 5~mm, enabling reliable matching of emulsion tracks to those reconstructed in BabyMIND. Additional emulsion films were attached to the upstream face of the ST, to further improve track matching between detectors. Further details of the ST are given in Ref.~\cite{odagawa_tracker}.

%% file: SubFiles/3_Simulation.tex
\section{Simulation}
\label{sec:simulation}

\subsection{Beam neutrino flux}

The neutrino flux prediction was evaluated following the standard T2K flux simulation framework \cite{t2k_flux}. It was based on the measured proton beam profile and horn currents during data taking. Interactions of the proton beam with the graphite target were modeled using \textsc{FLUKA}~2011.2x \cite{fluka_2005}. The production of secondary $\pi$ and $K$ mesons was tuned to data from the CERN NA61/SHINE replica target measurements \cite{NA61SHINE:2018rhe}. These secondary particles were then propagated with the \textsc{JNUBEAM} package \cite{t2k_flux}, based on \textsc{GEANT3} \cite{geant3}, and the neutrino flux at the detector was calculated from their decay kinematics.

\subsection{Neutrino interaction}

Neutrino-nucleus interactions were simulated using \textsc{NEUT}~5.6.4~\cite{neut}, with the same model configuration used in the T2K oscillation analysis~\cite{t2k_epjc_2023}. Relevant interaction channels for both charged-current (CC) and neutral-current (NC) processes in the $\mathcal{O}(0.1\text{--}10)$~GeV neutrino energy range were modeled. The simulated interaction channels included one-particle--one-hole (1p1h, corresponding to single-nucleon knockout) and two-particle--two-hole (2p2h) processes (for CC interactions only), single-pion production (1$\pi$) via nucleon resonances as well as non-resonant contributions and coherent scattering, and multi-pion production (Multi-$\pi$) with hadronic invariant mass $W$ below 2 GeV/$c^2$, and deep inelastic scattering (DIS, $W>2$ GeV/$c^2$) processes. Final-state interactions (FSI) of mesons and nucleons were modeled using a semi-classical intranuclear cascade. The model configuration and relevant inputs are summarized in Table~\ref{tab:neut_models}. 

Since some of the heavier nuclei present in the NINJA detectors (e.g., Ag and Br in the emulsion films) were not supported in NEUT, we supplied the nuclear charge density parameters required by the Nieves \textit{et al.} model~\cite{nieves, bourguille} for the 1p1h cross section calculations, with the numerical parameter values taken from Ref.~\cite{jager}. For other processes (e.g., 1$\pi$ production and FSI), we assumed a Fermi momentum of 250~MeV/$c$ and a nucleon binding energy of 35~MeV.

\begin{table*}[b]
\caption{Principal neutrino--nucleus interaction models implemented in \textsc{NEUT}~5.6.4~\cite{neut} and relevant to the present analysis. The configuration follows that adopted in the recent T2K oscillation analysis~\cite{t2k_epjc_2023}. FF and PDF denote form factors and parton distribution functions, respectively. Different 1p1h models are used for $^{12}\mathrm{C}$ and $^{16}\mathrm{O}$ than for the other target nuclei.}
\label{tab:neut_models}
\centering
\begin{tabular}{p{0.15\textwidth} p{0.4\textwidth} p{0.35\textwidth}}
\hline
\hline
Mode & Model & Major inputs and parameters \\
\hline
1p1h
&  Ground state: Benhar et al.~\cite{benhar, benhar_escat}
& \\
\cline{2-3}
($^{12}$C/$^{16}$O) & Cross section: Llewellyn-Smith ~\cite{llewellyn_smith}
& Vector FF: BBBA05 ~\cite{bbba05} \\
\cline{1-2}
1p1h& Nieves et al.~\cite{nieves, bourguille} 
& $M_A^{\mathrm{QE}} = 1.03~\mathrm{GeV}$  \\
(Others)& & \\
CC 2p2h& & \\
\hline
$1\pi$, Resonant
& Rein--Sehgal ~\cite{rein_sehgal, berger_sehgal}
& FF: Graczyk-Sobczyk \cite{graczyk}\\

& 
& $C_5^A(0)=1.06$ \\

&
& $M_A^{\mathrm{RES}}=0.91~\mathrm{GeV}$\\
\hline
1$\pi$, Coherent
& Berger--Sehgal ~\cite{bs_cohpi}
& \\
\hline
Multi-$\pi$ ($W<2$ GeV)
& Custom \cite{custom_pi}
& \\
\hline
DIS
& PYTHIA v5.72~\cite{pythia}
& PDF: GRV98 \cite{grv98} \\
($W>2$ GeV)
& 
    & with low-$q^2$ correction~\cite{bodek_yang} \\
\hline
FSI & Intranuclear cascade \cite{bertini} & $\pi$: Salcedo-Oset \cite{salcedo_oset} (tuned \cite{neut_pi_fsi_tune})\\
& & Nucleons: Bertini \cite{bertini_xsec}\\
\hline

\hline
\end{tabular}
\end{table*}

\subsection{Secondary interactions and detector response}
\label{sec:detector_response}

The NINJA detector geometry, including material composition and dimensions, was implemented in the existing WAGASCI--BabyMIND simulation framework \cite{wgbm_2025} based on Geant4 10.5.2 \cite{geant4, geant4_2, geant4_3}. Particle transport and secondary interactions in the detector and surrounding materials were modeled using the \texttt{QGSP\_BERT} physics list \cite{geant4_qgsp}.

Per-film tracking efficiencies and angular resolutions as functions of the incident track angle were measured using tracks traversing multiple adjacent films, predominantly cosmic-ray muons and ``sand muons'' (muons originating from neutrino interactions in material upstream of the detector, primarily the surrounding sand and rock). These measurements were implemented in the simulation as lookup tables. Figure~\ref{fig:film_eff} shows the measured per-film tracking efficiencies in the central ECC. The typical efficiency exceeds 98\% across most incident track angles. The slight angular dependence is primarily due to the longer path length of inclined tracks through the emulsion layer, resulting in a larger number of recorded silver grains.

The track angle and the volume pulse height (VPH; see Sec.~\ref{sec:partner_sel} for details), which is correlated with the energy loss per unit track length ($dE/dx$), are key observables measured in each emulsion film. These quantities are used for particle identification (PID) (Sec.~\ref{sec:partner_sel}) and momentum reconstruction (Sec.~\ref{sec:mcs_mom}). To reproduce the angular resolution observed in data, the simulated track angles in each film were smeared according to the measured angular spread as a function of track angle. The resulting angular information serves as the basis for momentum reconstruction. The VPH of each track was sampled from a likelihood function constructed using the VPH and $p\beta$ distributions observed in data. Details of the $p\beta$ variable and the likelihood construction are provided in Sec.~\ref{sec:mcs_mom}.

The track connection efficiencies across detector subsystems were evaluated using muons penetrating PM through BM. The BM--ST track connection efficiency was modeled directly in the Geant4-based detector simulation (overall 98.4\% in sand muon simulation; see Fig.~\ref{fig:bmst_eff} and Sec.~\ref{sec:bmst_matching} for validation). The ST--LES track connection efficiency measured using sand muon data (overall 86.7\%; see Fig.~\ref{fig:bm_les_ecc_eff} and Sec.~\ref{sec:mu_connection} for the evaluation procedure) was implemented in the simulation as a track-angle-dependent lookup table, which was applied as an event weight to all simulated muon tracks. These muon connection efficiencies were considered in the systematic uncertainty evaluation (see Sec.~\ref{sec:syst_mueff}).

\Fig{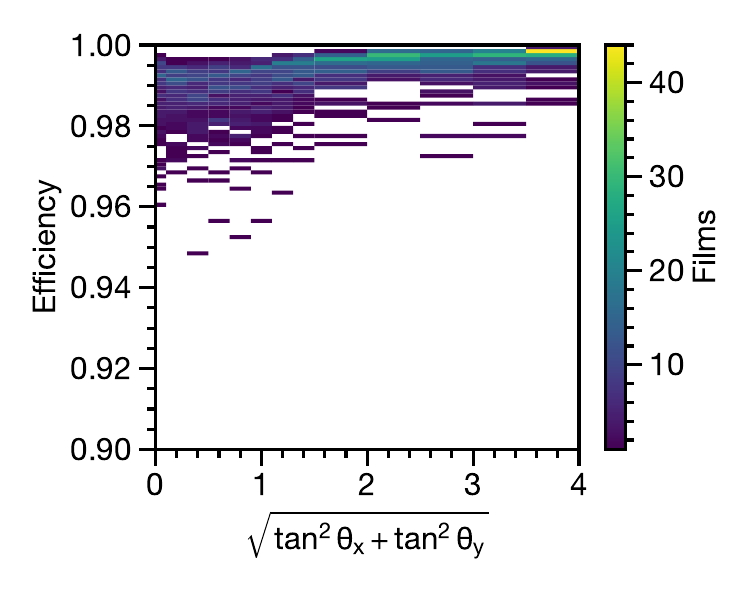}{0.5\textwidth}{Distribution of tracking efficiencies for the 118 films spanning the central ECC water volume, shown as a function of the incident track angle. Here, $\theta_x$ and $\theta_y$ denote the track angles with respect to the film normal in the horizontal ($x$) and vertical ($y$) directions, respectively, as illustrated in Fig.~\ref{fig:detectors}.}{fig:film_eff}

%% file: SubFiles/4_Analysis.tex
\section{Particle identification and reconstruction}

\subsection{Track recognition in emulsion detectors}
\label{sec:track_recognition}

After beam exposure, all films were chemically developed and scanned with the Hyper Track Selector (HTS) \cite{hts}, a high-speed automated optical microscope system optimized for large-area emulsion readout. Each emulsion layer was scanned in depth over a total thickness of 60~µm (after development shrinkage). The resulting tomographic images were binarized to reduce data volume and enable efficient pattern recognition.

Linearly aligned silver grains in the tomographic images were reconstructed as ``microtracks'' \cite{aoki_ts, netscan}, representing track segments within a single emulsion layer on each side of the polystyrene base. The microtrack recognition algorithm has been extended to large-angle tracks \cite{suzuki_large_angle}, enabling reliable reconstruction of fully connected particle trajectories for tracks satisfying $|\tan\theta|<4.0$ ($\theta \in [0^\circ,76^\circ)\cup(104^\circ,180^\circ]$), where $\theta$ denotes the track incident angle with respect to the emulsion film normal. A pair of microtracks, each reconstructed on opposite sides of the polystyrene base, was matched to form a ``basetrack'' by requiring consistency in position and angle, with corrections applied for film distortions and emulsion shrinkage during development.

A global alignment was first performed to correct for overall translations and rotations of each film, followed by local alignment to account for position-dependent distortions across the film surface. The alignment parameters were determined using reference through-going tracks and were iteratively optimized by minimizing position and angular residuals between consecutive films, achieving $O(1)$ µm precision in their relative positioning.

After alignment, basetracks in successive films were connected to reconstruct physical particle trajectories. Candidate connections were identified by extrapolating basetracks to adjacent films and searching for compatible matches within predefined position and angular tolerances. The most probable connections were selected by requiring continuity of track parameters, including position, angle, and track quality indicators such as grain density. Further details of the emulsion track recognition process are given in Refs.~\cite{suzuki_large_angle, suzuki_thesis}.

\subsection{Effective momentum \texorpdfstring{$p\beta$}{p-beta} reconstruction based on multiple Coulomb scattering}

\label{sec:mcs_mom}

The momenta of charged particle tracks in the ECC were reconstructed from the angular fluctuations of track segments measured in successive emulsion films, arising from MCS primarily in the stainless steel plates placed between adjacent films. According to the Highland formula~\cite{highland,highland2}, the quantity $p\beta$, where $p$ is the particle momentum and $\beta=v/c$, is inversely proportional to the root-mean-square width of the MCS angular distribution. The value of $p\beta$ and its uncertainty were estimated using a likelihood fit to the observed scattering angles along the track, taking into account energy loss in the detector materials between adjacent basetracks~\cite{odagawa_momentum}. This method is applicable to muons with momenta up to approximately $1~\mathrm{GeV}/c$ and achieves a momentum resolution of about 10--30\%~\cite{odagawa_momentum}.

\subsection{Identification of emulsion tracks coincident with beam spills}
\label{sec:mu_connection}

Beam-related tracks in the ECC were identified by starting from reconstructed BM tracks associated with a beam spill and propagating the track matching upstream through the ST, LES, and ECC (for the BM reconstruction procedure, see Ref.~\cite{wgbm_2025}). Once a BM track was matched to an ECC track, all tracks connected to the corresponding interaction vertex were assigned to the same neutrino event.

\subsubsection{BM--ST track matching}
\label{sec:bmst_matching}

First, the BM track was linearly extrapolated in the bending (vertical, $y$) direction using hits in the three most upstream scintillator layers to reduce the effect of magnetic-field-induced bending, while all reconstructed hits were used for the linear fit in the non-bending (horizontal, $x$) direction. Matching ST hits were then searched for within position tolerances of 30~cm in $x$ and 20~cm in $y$ directions, following Ref.~\cite{odagawa_tracker}, reflecting the asymmetric BabyMIND hit position resolutions in the $x$ and $y$ directions.

\begin{figure}[htb]
  \centering
  \begin{minipage}[t]{0.48\textwidth}
    \centering
    \includegraphics[width=\linewidth]{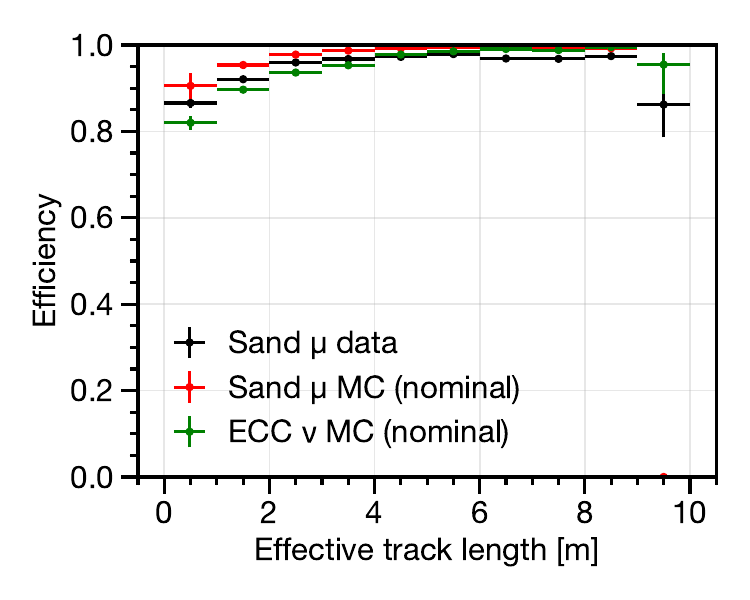}
    
  \end{minipage}
  \hfill
  \begin{minipage}[t]{0.48\textwidth}
    \centering
    \includegraphics[width=\linewidth]{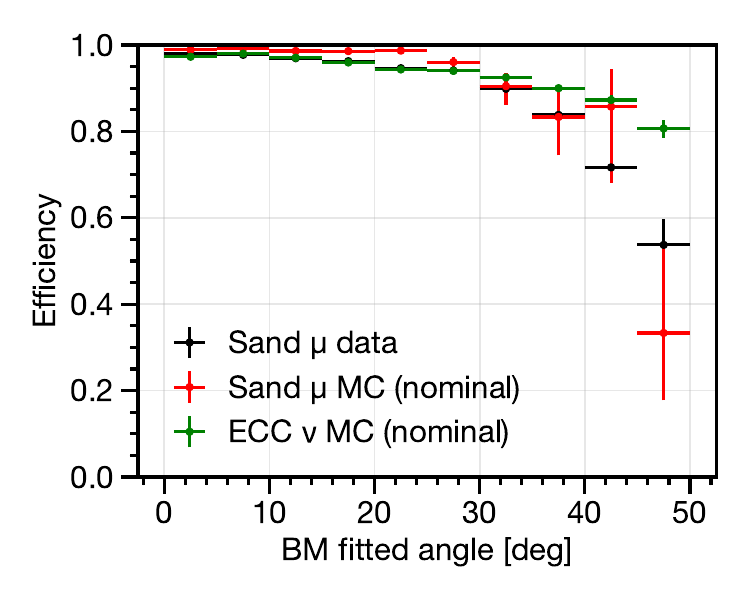}

  \end{minipage}
\caption{BM--ST track-connection efficiency as a function of effective track range (left) and estimated track incident angle at BM (right). Black points represent through-going muon data, dominated by sand muons. Red points represent sand muon MC with the same upstream detector selection requiring PM hits. For reference, green points show the efficiency for muons produced in neutrino interactions within the ECC, without the PM hit requirement.}
  \label{fig:bmst_eff}
\end{figure}

Figure~\ref{fig:bmst_eff} shows the BM--ST track-connection efficiency as a function of effective track range and track angle for through-going muons. The effective track range is defined as the sum of the path lengths traversed in each material, weighted by their continuous slowing down approximation (CSDA) stopping powers relative to iron. The efficiency, defined as the fraction of BM tracks successfully matched to an ST hit, was evaluated using data with hits in the upstream Proton Module (PM), together with MC samples of sand muons. To ensure that the tracks traverse the sensitive area of the ST, only BM tracks extrapolating to its central region were included. The overall BM--ST track-connection efficiency was 96.2\% in the sand muon data and 98.4\% in the sand muon simulation.

\subsubsection{ST--LES--ECC track matching}

After establishing the BM--ST connection, the corresponding LES track was searched for using the track timestamp. Initial position and angular consistency requirements between the BM--ST track and LES tracks were applied to identify candidate tracks. A track-matching $\chi^2$ was then calculated from the position and angular residuals between successive pairs of tracks along the beam direction, comprising the BM--ST track and those recorded in the stationary films of the LES, with each residual normalized by the corresponding detector resolution. Loose requirements on these $\chi^2$ values were applied to maximize signal efficiency while suppressing accidental matches.

Figure~\ref{fig:bm_les_ecc_eff} shows the probability of finding an LES track with the matching timestamp within the $\chi^2$ selection. The probability of an accidental LES track passing the same selection was estimated from a control sample obtained by searching for LES tracks with timestamps shifted by 12 hours. Assuming statistical independence between the presence of a true matching track and an accidental background track, the background-corrected efficiency of finding the true LES track associated with the BM--ST track was derived (overall 86.7\%).

If multiple LES tracks survived the selection, all were propagated to search for a matching ECC track, where the connection efficiency was found to be nearly 100\% and was therefore assumed to be unity. Approximately 95\% were matched to ECC-penetrating tracks (e.g., sand muons or cosmic muons) which would be rejected at the later muon selection stage (Sec.~\ref{sec:mu_sel}). If multiple LES--ECC candidates with vertices reconstructed inside the ECC remained, the candidate with the smallest overall matching $\chi^2$ was selected. The probability that an accidental LES track would have a smaller $\chi^2$ than the true track was estimated to be 1.1\% using a bootstrap method based on the true-match and background $\chi^2$ distributions extracted from the matching-timestamp sample and the 12-hour-shifted control sample. The resulting track misconnection probability for the selected muon sample with vertices inside the ECC was estimated to be of order $0.1\%$ or less, and was therefore neglected in this analysis. Further details of the track matching procedure across detector subsystems are given in Refs.~\cite{odagawa_tracker, odagawa_thesis}.

\Fig{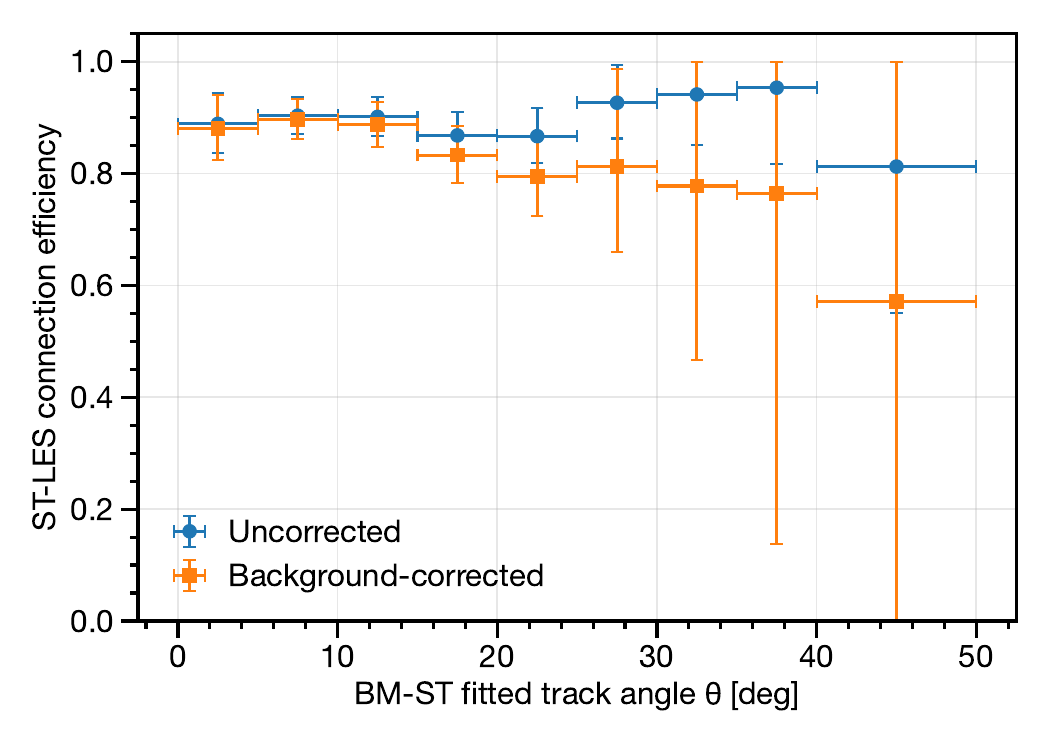}{0.7\textwidth}{ST--LES track connection efficiency in data as a function of the BM--ST track angle. The BM--ST track angle was obtained from a linear fit to the BM and ST hit positions. Blue points show the uncorrected efficiency, while orange points show the background-corrected efficiency. Error bars are derived from toy MC simulations assuming Poisson counting statistics.}{fig:bm_les_ecc_eff}

\subsection{Muon selection}
\label{sec:mu_sel}

Among all BM tracks connected to the central ECC, the longest track in each beam bunch was identified as the muon candidate, resulting in a total of \textcolor{black}{9,371} candidate tracks. Most of these were through-going sand muons originating from neutrino interactions in materials upstream of the ECC. Of these, \textcolor{black}{386} events had an interaction vertex reconstructed within the ECC (a $22\times22$~cm$^2$ film area with an approximately uniform event rate). The fiducial area of each emulsion film was defined as the central $17 \times 17$~cm$^2$ region, providing sufficient margins for reconstructing hadron tracks emitted at large angles. After applying the fiducial volume requirement, 227 events remained. The muon candidate purity was estimated to be 97\% from the MC simulation.

The interaction material of each selected event was then identified by visual inspection under a microscope. The upstream endpoint of the muon candidate track was located with sub-\textmu m precision. If the most upstream microtrack penetrated the emulsion layer, the interaction was assigned to the first upstream material layer; otherwise, it was classified as occurring in the emulsion film. Interactions occurring in the envelope surrounding the emulsion films cannot be distinguished from those occurring in the adjacent water layers (see Fig.~\ref{fig:ecc}). An estimated 6.6 interactions in the envelope were predicted to remain as background in the final sample. In the following, the upstream endpoint of the muon microtrack recorded in each film was assumed to be identified without ambiguity.

Table~\ref{tab:ecc_material} summarizes the material composition of the central ECC fiducial volume together with the observed and expected numbers of selected neutrino interactions in each material. Table~\ref{tab:event_selection} summarizes the event selection and the corresponding numbers of selected events compared with the nominal MC prediction. The observed interaction rates in the stainless steel plates and emulsion layers differ from the MC prediction, potentially reflecting imperfect modeling of neutrino interactions on heavy non-isoscalar nuclei, such as Fe and Cr in the stainless steel and Ag and Br in the nuclear emulsion. To conservatively account for this effect, the 18\% discrepancy between the observed and predicted numbers of interactions reconstructed within the fiducial volume was assigned as the systematic uncertainty associated with background contamination (see Sec.~\ref{sec:syst_bg}).

\begin{table}[tb]
\centering

\caption{Material composition of the fiducial volume in the central ECC water target section. The ``Data'' and ``ECC MC'' columns show the observed and predicted numbers of neutrino interactions in each material. The ``ECC MC'' column includes only interactions occurring within the ECC and is categorized according to the true interaction material. The first-row ``ECC MC'' entry (79.3) includes 6.6 events originating in the tracking unit envelope.}

\label{tab:ecc_material}
\begin{tabular}{lrrr}
\hline
\hline
Material & Mass [kg] & Data [events] & ECC MC [events] \\
\hline

Water & 3.86 & \multirow{2}{*}{82} &  \multirow{2}{*}{81.5} \\
\cline{1-2}
Tracking unit envelope & 0.42 & & \\
\hline
Stainless steel (SUS316L) & 6.80 & 115 & 146.8 \\
\hline
Emulsion gel & 1.46 & \multirow{2}{*}{30} & \multirow{2}{*}{40.8} \\
\cline{1-2}
Polystyrene (film base) & 0.86 &  & \\

\hline
Total & 13.40 & 227 & 269.1 \\
\hline 
\hline
\end{tabular}
\end{table}

\begin{table}[tb]
\centering 
\caption{Summary of the muon selection criteria and the corresponding event yields in data and MC simulation. ``ECC'' indicates interactions in the central ECC, while ``Other'' represents those in all other surrounding material. ``$\nu_\mu$ CC on water'' represents the expected signal sample size, and ``Else'' includes all remaining ECC interactions. The dominant contribution to the ``Else'' category is neutrino interactions in the tracking unit envelope, of which 6.6 events are expected to remain after the final selection.} \label{tab:event_selection}

\begin{tabular}{lrrrrr} 
\hline 
\hline 
\multirow{3}{*}{Selection step} & \multirow{3}{*}{Data} & \multicolumn{4}{c}{MC} \\ 
\cline{3-6} & & \multirow{2}{*}{Total} & \multicolumn{2}{c}{ECC} & \multirow{2}{*}{Other} \\ 
\cline{4-5} & & & $\nu_\mu$ CC on water & Else &
\\ 
\hline 
BM-connected $\mu$ candidates & 9,371 & 10,402.9 & 144.3 & 461.2 & 9,797.4 \\ 
Vertex within ECC target section & 386 & 435.6 & 108.5 & 309.9 & 17.2 \\ 
Vertex within fiducial volume & 227 & 277.0 & 70.4 & 198.7 & 7.9 \\ 
Vertex identified in water layers & 82 & 87.5 & 70.4 & 11.1 & 6.1 \\ 
\hline \hline
\end{tabular} 
\end{table}

\subsection{Hadron identification and selection}
\label{sec:partner_sel}

Hadron candidates associated with the muon vertex were identified by selecting tracks that penetrated at least two films and whose minimum distance from the extrapolated muon track was below a variable threshold, typically less than 0.3~mm. The threshold depended on the muon angle and momentum, accounting for the track angular resolution and MCS in the water layer \cite{suzuki_thesis}. The effective momentum $p\beta$ of hadron tracks was reconstructed from MCS angles in the same manner as for muons, as described in Sec.~\ref{sec:mcs_mom}.

\subsubsection{Likelihood construction for proton and pion PID}

Particle identification was performed by comparing the likelihood of each track being either a proton or a charged pion. The primary discriminating variable was the volume pulse height (VPH) \cite{toshito_vph}, which is strongly correlated to the number of developed silver grains along a track and is therefore closely related to the ionization energy loss, $dE/dx$. A correction for the gradual time dependence of the film response was applied using timing information provided by muon tracks.

\Fig{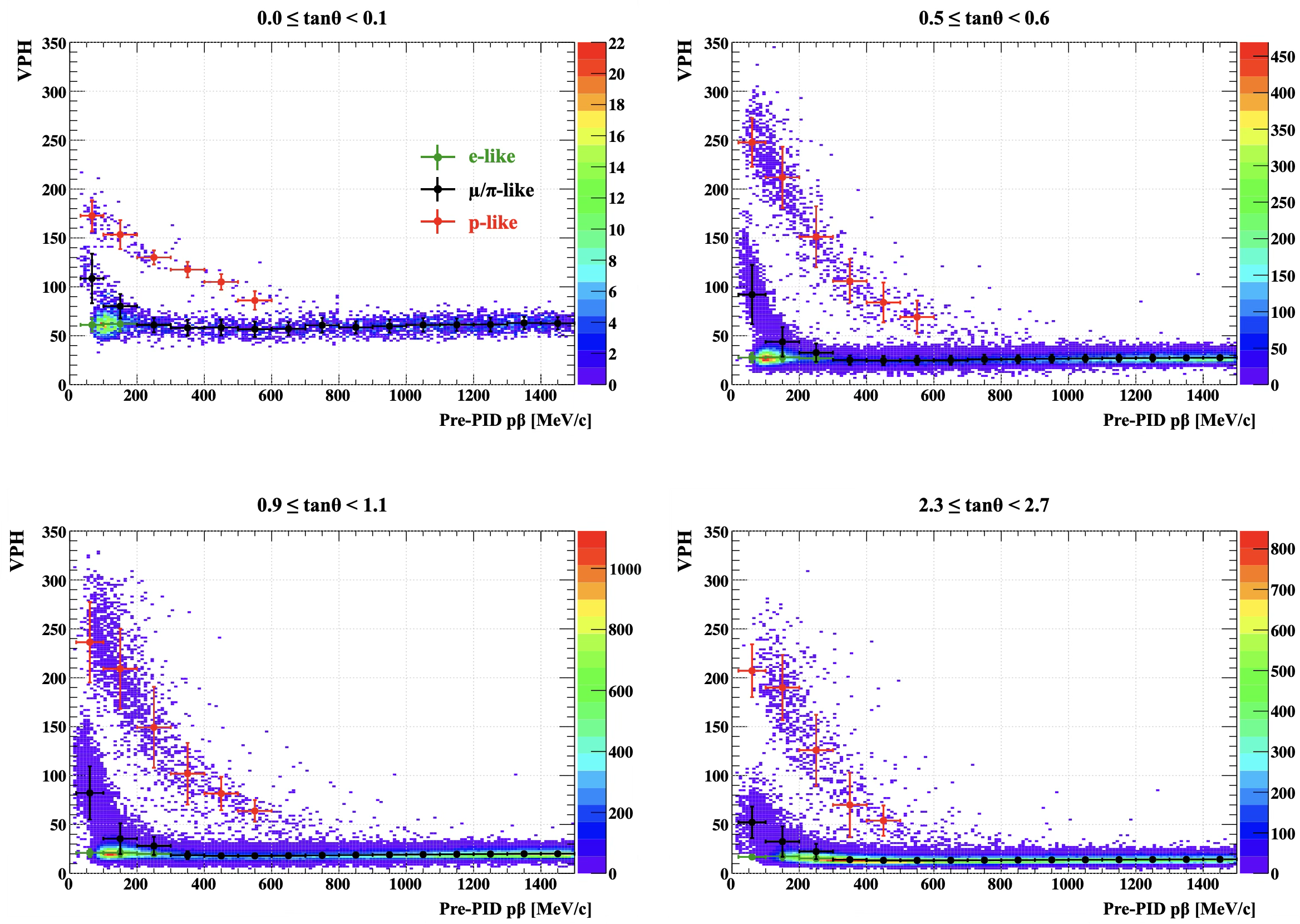}{\textwidth}{Two-dimensional distributions of VPH versus pre-PID $p\beta$ for basetracks in the central ECC, shown for different track angle regions, where $\theta$ denotes the track's incident angle on each film. The panels range from forward-going tracks (top left) to larger-angle tracks (bottom right). Gaussian fits were performed in each $p\beta$ bin of width 100~MeV/$c$, using separate VPH peak regions corresponding to different particle hypotheses: electron (green), muon/pion (black), and proton (red).}{fig:pid}

The proton and pion likelihoods were constructed from probability density functions (PDFs) parameterized in bins of track angle and ``pre-PID'' $p\beta$, where $p\beta$ was reconstructed assuming the energy loss of a muon. The PDFs were derived primarily from the VPH distributions of basetracks observed in data. Figure~\ref{fig:pid} shows the two-dimensional distributions of VPH versus pre-PID $p\beta$ for several track angle regions. In each track-angle and $p\beta$ bin, the observed VPH distribution was fitted with up to three Gaussian components corresponding to electrons, muons/pions, and protons, ordered by increasing VPH mean, as illustrated in Fig.~\ref{fig:pid_pdf_fit}. Protons generally produce tracks with larger VPH values compared to charged pion tracks at the same pre-PID $p\beta$.

\Fig{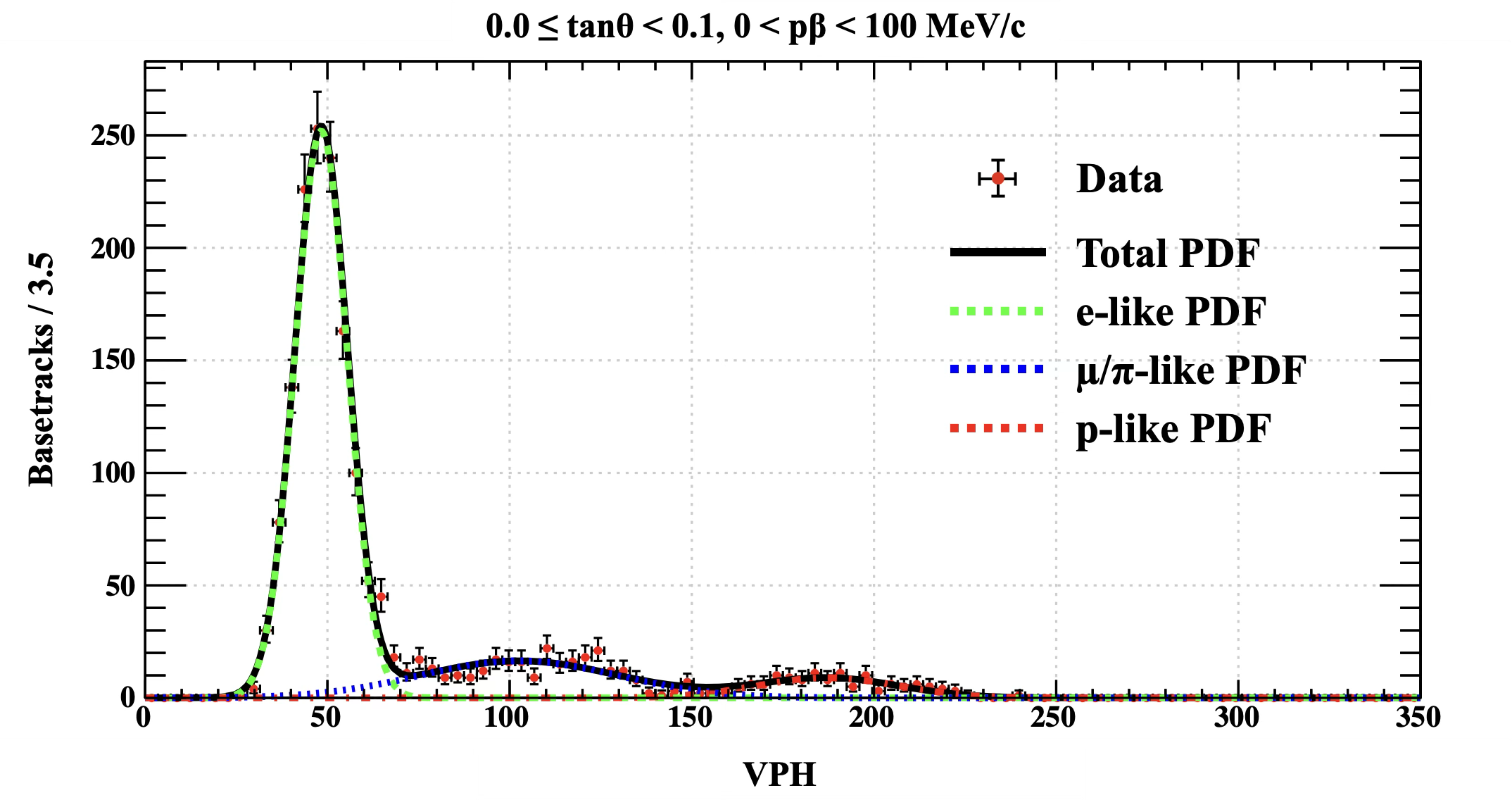}{0.8\textwidth}{VPH distribution of basetracks associated with reconstructed tracks in the region $0<p\beta<100$~MeV/$c$ and $0<\tan\theta<0.1$, where $\theta$ is the track incident angle with respect to the film normal. The observed distribution is fitted with the sum of three Gaussian PDFs representing electron-like (green; likely originating from cosmic-ray muon interactions), muon/pion-like (blue), and proton-like (red) tracks.}{fig:pid_pdf_fit}

Muons and charged pions behave approximately as minimum ionizing particles (MIPs) for pre-PID $p\beta>200$~MeV/$c$, while the proton ionization energy loss approaches the MIP value for $p\beta>700$~MeV/$c$, degrading the VPH-based PID separation. To extend the PID capability to this region, where stable multi-Gaussian fits often failed, the Gaussian parameters for $p\beta\in[200,700]$~MeV/$c$ were extrapolated from bins with successful multi-Gaussian fits. The proton-like VPH means were extrapolated as a linear function of the Bethe--Bloch ionization energy loss, while the Gaussian widths were extrapolated as a linear function of $\sqrt{\mathrm{VPH}}$, motivated by the Poisson statistics of the developed silver grains.

In the low-$p\beta$ region ($p\beta\lesssim200$~MeV/$c$), where the VPH response begins to saturate with increasing ionization energy loss, the Gaussian means were instead obtained directly from the data through Gaussian fits and interpolation. Furthermore, in the region where the muon and charged pion ionization energy loss reaches a minimum ($p\beta\in[200,400]$~MeV/$c$), the VPH distributions become asymmetric. Therefore, empirical non-Gaussian PDFs derived directly from the observed VPH distributions were used instead of Gaussian parameterizations.

The proton and pion likelihoods, $\mathcal{L}_\mathrm{p}$ and $\mathcal{L}_\mathrm{\pi}$, were calculated from the cumulative probabilities of the corresponding PDFs evaluated at the observed VPH. The resulting likelihood ratio, $\mathcal{L}_\mathrm{\pi}/(\mathcal{L}_\mathrm{\pi}+\mathcal{L}_\mathrm{p})$, is shown in Fig.~\ref{fig:pidll}, demonstrating good agreement between data and simulation. Tracks with a likelihood ratio below 0.5 were classified as protons, while the remainder were classified as charged pions. Particle identification was applied only to tracks with $p\beta<700$~MeV/$c$, where protons can be reliably distinguished from MIPs. Tracks with $p\beta\geq700$~MeV/$c$ (``MIP-like'') were therefore not assigned a particle identity and contribute only to the total charged track multiplicity, but not to particle-specific observables such as the proton and pion kinematic distributions.

\Fig{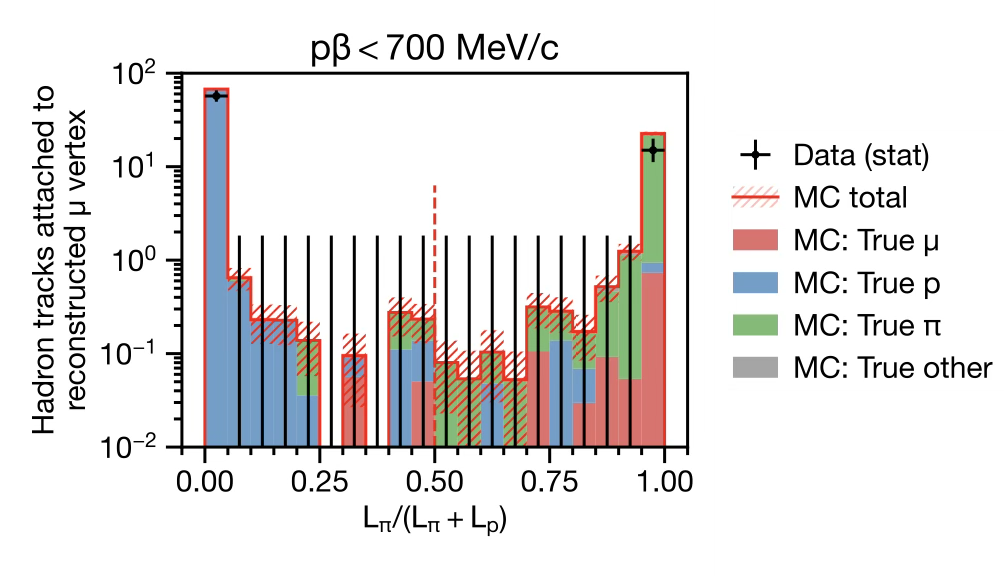}{0.7\textwidth}{Distribution of the likelihood ratio $\mathcal{L}_{\pi}/(\mathcal{L}_{\pi}+\mathcal{L}_\mathrm{p})$ for hadron tracks associated with the selected muons, comparing data with MC simulation. Tracks with $p\beta<700$~MeV/$c$ are shown. The red dashed line indicates the proton/pion classification threshold at 0.5. The hatched uncertainty band shows the statistical uncertainty of the MC simulation.}{fig:pidll}

\subsubsection{Background rejection}

Background tracks were removed through several criteria. Pion-like and MIP-like tracks containing fewer than 10 basetracks were rejected to suppress contamination from accidental linear connections of basetracks, whose occurrence decreases exponentially with track length. Tracks identified by microscope inspection as entering background particles, predominantly cosmic-ray muons, were rejected. Tracks that could be connected under specific film alignments, likely accumulated during film transportation, were also rejected. In addition, contained tracks with ranges exceeding those expected from their MCS-based momentum estimates were rejected. In events with more than three hadron candidates, tracks associated with a secondary vertex were additionally removed. The main interaction vertex was identified as the vertex with the larger number of associated candidates and the better track convergence.

In total, 79 tracks were identified as hadrons: 57 were classified as protons, 15 as pions, and the remaining 7 were left unclassified because their reconstructed $p\beta$ exceeded 700~MeV/$c$. Figure~\ref{fig:partner_eff} shows the estimated track selection efficiency, defined as the fraction of particle tracks that are correctly identified as the corresponding particle species, in the MC simulation as a function of particle kinematics.

\Fig{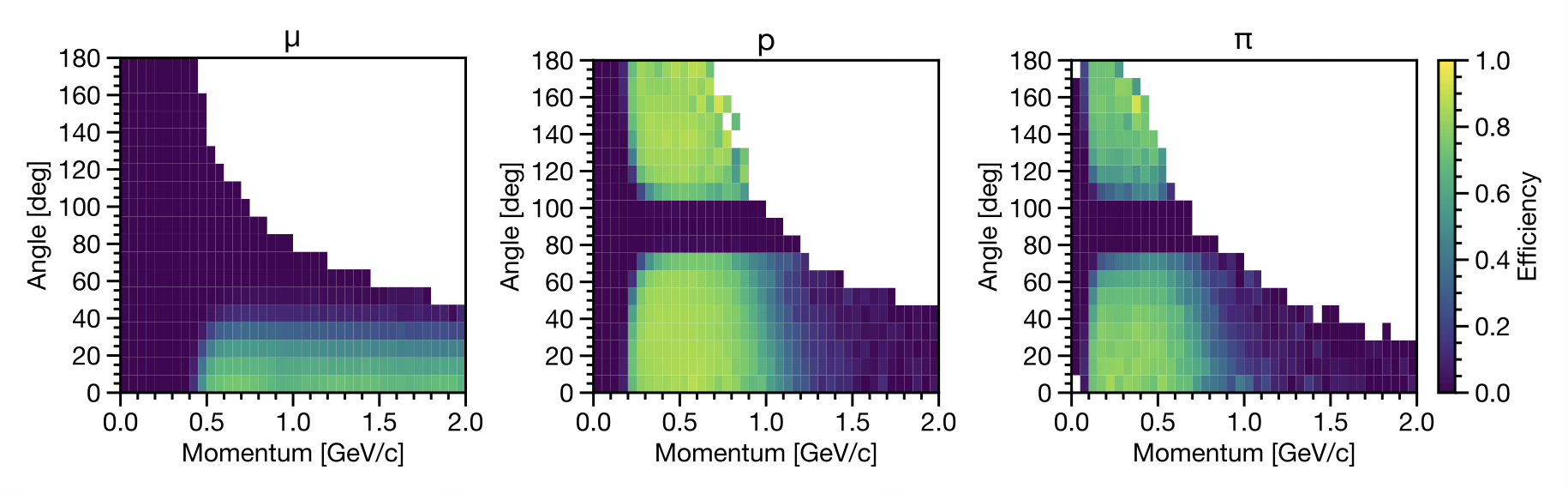}{\textwidth}{Track selection efficiency evaluated using MC simulation as a function of the true particle angle and momentum for muons (left), protons (center), and pions (right). Only bins containing 10 or more MC tracks are displayed.}{fig:partner_eff}

\subsubsection{Performance and validation of MCS-based proton momentum reconstruction}
\label{sec:p_mom_vs_len}

Figure~\ref{fig:pmomres} shows the expected momentum reconstruction performance for proton-like tracks in simulation, with a resolution of approximately 10--20\%. The reconstruction was further validated by comparing the correlation between the reconstructed momentum and track length in the ECC between data and MC, as shown in Fig.~\ref{fig:p_2d_corr}. The MC correlation shape was modeled using a Gaussian kernel density estimate (KDE), and the data--MC agreement was quantified with a likelihood-based $p$-value obtained from bootstrap pseudo-experiments. Specifically, MC samples with the same size as the data were repeatedly resampled, and the $p$-value was defined as the probability of obtaining a KDE log-likelihood smaller than that observed in the data.

\Fig{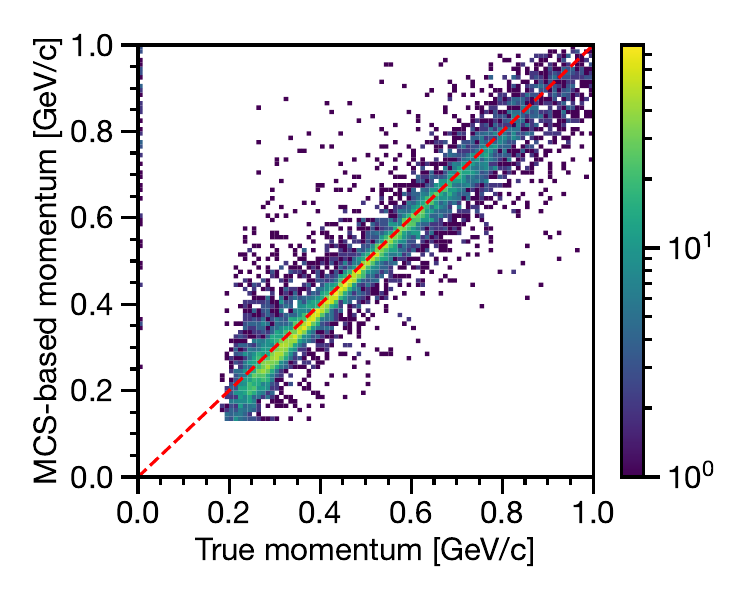}{0.5\textwidth}{True momentum versus MCS-reconstructed momentum in MC for selected proton tracks in the final sample.}{fig:pmomres}

While the one-dimensional distributions of reconstructed momentum and track length are consistent with the data within statistical uncertainties, the two-dimensional shape exhibits a modest discrepancy, yielding a $p$-value of approximately 5\%, largely independent of the KDE bandwidth. The discrepancy may originate from a small number of outliers, such as tracks with short reconstructed lengths but large reconstructed momenta. Such tracks could arise from hadron scattering in the detector. In the current reconstruction, strongly scattered tracks may have shortened reconstructed lengths because only approximately linear sequences of basetracks are connected. In addition, the reconstructed momentum is sensitive to the basetrack angular resolution, and mismodeling of the angular resolution may affect the reconstructed momentum distribution. The impacts of the hadron scattering model and the basetrack angular resolution are treated as systematic uncertainties, as discussed in Sec.~\ref{sec:syst_det}.

\Fig{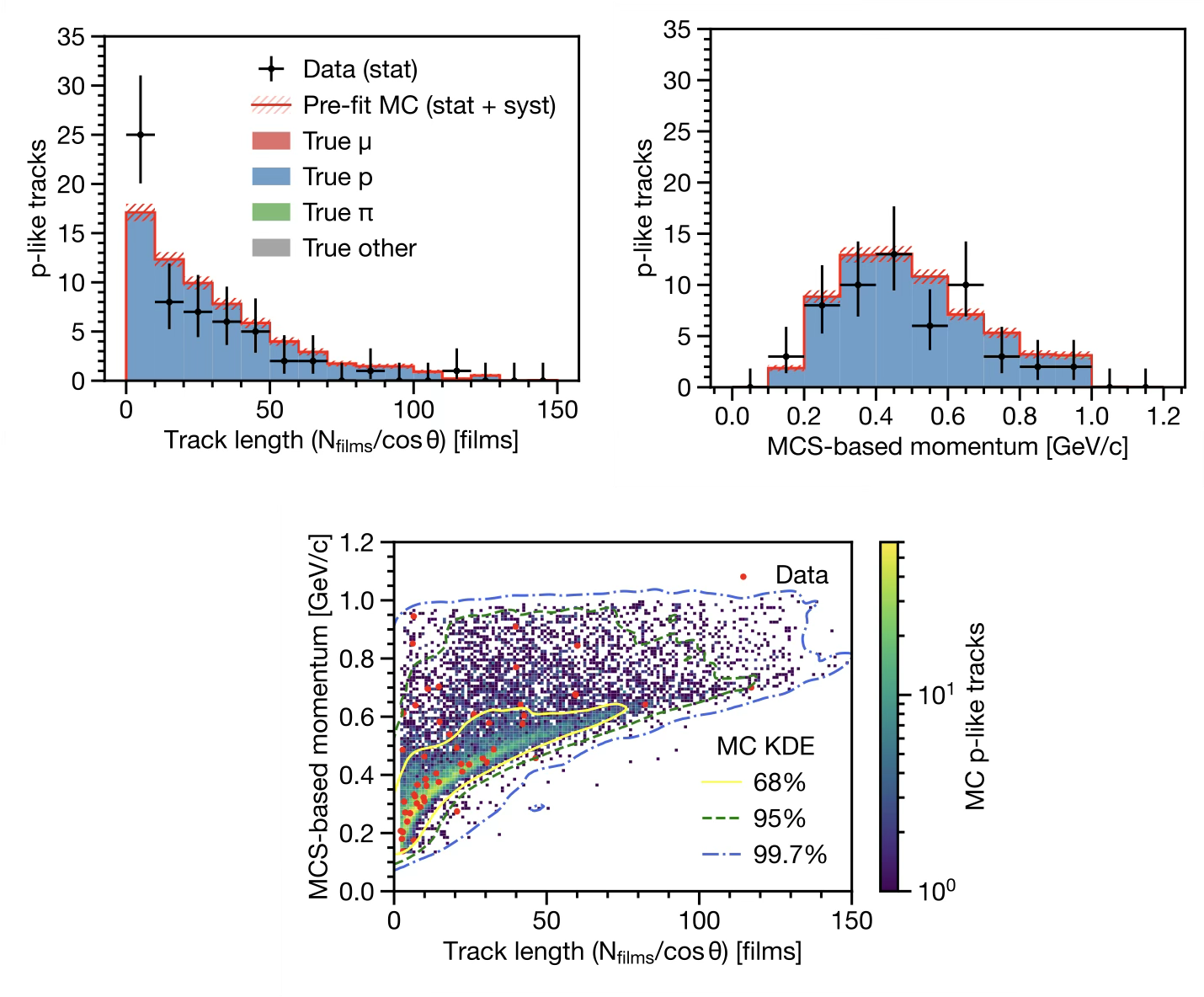}{\textwidth}{Distributions of the track length proxy, defined as the number of emulsion films traversed ($N_\text{films}$) divided by $\cos\theta$, where $\theta$ is the track incident angle, and the momentum reconstructed from MCS for proton-like tracks. The top panels show the one-dimensional projections, while the bottom panel shows the corresponding two-dimensional distribution. The red hatched bands in the top panels indicate the MC statistical uncertainty. The colored contours in the bottom panel show the 68\%, 95\%, and 99.7\% probability density regions obtained from a Gaussian KDE of the MC sample.}{fig:p_2d_corr}

%% file: SubFiles/5_Systematics.tex
\section{Systematic uncertainties}
\label{sec:syst}

In the following, we briefly describe the sources of systematic uncertainty and their corresponding bin-by-bin fractional uncertainties for each observable. The effects of neutrino interaction and detector response uncertainties were evaluated using covariance matrices adopted from the systematic uncertainty studies presented in Ref.~\cite{odagawa_thesis}. 

\subsection{Covariance matrix construction}

For each systematic source, the uncertainty in a reconstructed observable was evaluated using the corresponding upward and downward variations, representing $+1\sigma$ and $-1\sigma$ shifts of the underlying systematic parameter. Let $N_i$ denote the nominal MC prediction in bin $i$, and $N_i^{+}$ and $N_i^{-}$ the predictions obtained from the $+1\sigma$ and $-1\sigma$ variations, respectively. The covariance matrix contribution from that source was constructed as

\begin{equation}
C_{ij}
\equiv
\frac{1}{2}
\left[
(N_i^{+}-N_i)(N_j^{+}-N_j)
+
(N_i^{-}-N_i)(N_j^{-}-N_j)
\right].
\label{eq:covmat}
\end{equation}

The corresponding fractional covariance matrix is defined as $F_{ij}\equiv C_{ij}/N_iN_j$. This construction incorporates the bin-to-bin correlations induced by the systematic variation and allows for asymmetric upward and downward shifts relative to the nominal prediction. The covariance matrices from the individual systematic sources were summed to obtain the total covariance matrix, assuming the sources are independent. For the goodness-of-fit evaluation (Sec.~\ref{sec:gof}), the resulting covariance matrix was represented in terms of independent Gaussian-constrained nuisance parameters obtained from its eigenvalue decomposition.

\subsection{Sources of systematic uncertainty}
\subsubsection{Beam neutrino flux}

The total uncertainty in the neutrino flux was estimated to be approximately 5--6\% around the peak neutrino energy. The dominant contributions arise from two main sources: uncertainties in hadron production from hadronic interactions within the graphite target, which are largely constrained by the NA61/SHINE measurements using a replica target at the same proton beam energy \cite{na61_replica}, and uncertainties in beamline modeling, including proton beam intensity, profile, magnetic horn, and surrounding materials and geometry. 

The flux uncertainty covariance matrix in true neutrino energy bins was adopted from the WAGASCI-BabyMIND analysis~\cite{wgbm_2025}, which uses the same beamline configuration and is located at almost the same off-axis position as the NINJA detector. Figure~\ref{fig:fluxunc} shows the fractional uncertainties from various sources as a function of true neutrino energy. The covariance matrix in true neutrino energy, $\mathbf{C}_{\textrm{flux}}$, was propagated to the reconstructed observable space using a detector response matrix, $\mathbf{R}$, constructed from the MC-predicted migration between true neutrino energy and the reconstructed observable. The covariance matrix in reconstructed space due to flux uncertainties was then given by $\mathbf{C}_{\mathrm{reco}}
=
\mathbf{R}\mathbf{C}_{\textrm{flux}}\mathbf{R}^{T}$. The resulting uncertainty in the reconstructed observables is typically at the level of 5--10\%, depending on the observable and bin.

\Fig{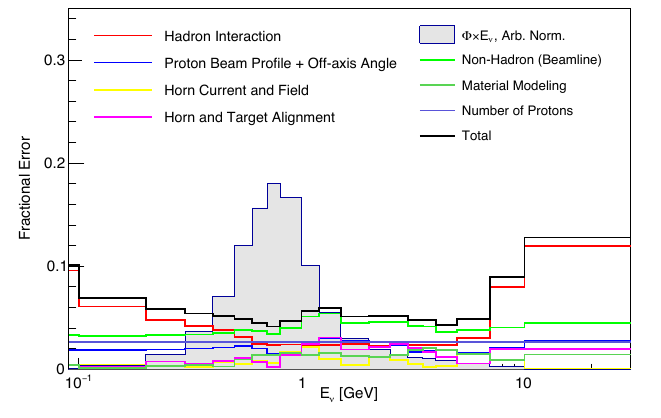}{0.7\textwidth}{Fractional uncertainties from different sources associated with neutrino flux modeling at the detector location (in $\nu_\mu$ beam mode), as a function of neutrino energy. The gray histogram shows the expected $\nu_\mu$ flux shape (arbitrary units). Reprinted from Ref.~\cite{odagawa_thesis}.}{fig:fluxunc}

\subsubsection{Neutrino interaction modeling}

To evaluate the impact of neutrino interaction model uncertainties, simulated events were reweighted according to variations of the relevant interaction model parameters. Each parameter was varied independently within its estimated $1\sigma$ uncertainty under a Gaussian constraint, as summarized in Table~\ref{tab:model_params}. The response of the predicted event distributions to these parameter variations was assumed to be approximately linear within the considered uncertainty range. Parameter correlations were neglected. The resulting uncertainty from neutrino interaction modeling varies between 10\% and 30\%, depending on the observable bin.

\begin{table}[htbp]
\centering
\caption{Summary of the model parameters, their central values, and corresponding 1$\sigma$ uncertainties used in the covariance matrix calculation defined in Eq.~\ref{eq:covmat}, adapted from Ref.~\cite{odagawa_thesis}. Here, $Q^2$ denotes the four-momentum transfer squared in units of GeV$^2/c^2$. SF refers to the nucleon spectral function model for oxygen~\cite{benhar, benhar_escat}, while MEC denotes meson exchange currents for oxygen, one of the dominant contributions to 2p2h interactions together with correlated nucleons. The parameters listed here represent a subset of those used in T2K publications~\cite{t2k_epjc_2023, wgbm_2025}.}
\label{tab:model_params}
\begin{threeparttable}
\begin{tabular}{lcc}
\hline
\hline
Parameter & Central value & $1\sigma$ uncertainty \\
\hline
CC 1p1h axial mass $M_A^{\mathrm{QE}}$ & $1.03~\mathrm{GeV}/c^2$ & $0.196~\mathrm{GeV}/c^2$ \\
CC 1p1h cross section normalization ($0.25 < Q^2 < 0.5$) & $1$ & $0.11$ \\
CC 1p1h cross section normalization ($0.5 < Q^2 < 1$) & $1$ & $0.18$ \\
CC 1p1h cross section normalization ($Q^2 > 1$) & $1$ & $0.40$ \\
SF optical potential correction scale & $0$ & $0.19$ \\
\hline
2p2h cross section normalization & $1$ & $0.20$ \\
2p2h cross section shape (target being $np$ or $nn$ pair) & $0$ & $0.33$ \\
2p2h cross section shape (MEC fraction for $np$ target) & $0$ & $0.20$ \\
2p2h cross section shape (MEC fraction for $nn$ target) & $0$ & $0.20$ \\
\hline
Effective binding energy for resonant single-$\pi$ production  & $25~\mathrm{MeV}$ & $5~\mathrm{MeV}$ \\
Isospin-$1/2$ non-resonant backgrounds & $1.3$ & $0.15$ \\
Resonant axial transition scale $C_A^{5}(0)$ & \textcolor{black}{$1.06$}\tnote{a} & $0.15$ \\
CC resonant axial mass $M_A^{\mathrm{RES}}$ & \textcolor{black}{0.95$~\mathrm{GeV}/c^2$}\tnote{a} & $0.15~\mathrm{GeV}/c^2$ \\
\hline
DIS/Multi-$\pi$ cross section normalization & 1 & 0.1\\
\hline
FSI: $\pi$ absorption & 1.404 & 0.432 \\
FSI: $\pi$ charge exchange ($>500$ MeV/$c$) & 1.8 & 0.288 \\
FSI: $\pi$ charge exchange ($<500$ MeV/$c$) & 0.697 & 0.305 \\
FSI: $\pi$-induced hadron production & 1.002 & 1.101 \\
FSI: $\pi$ QE scattering ($>500$ MeV/$c$) & 1.824 & 0.859 \\
FSI: $\pi$ QE scattering ($<500$ MeV/$c$) & 1.069 & 0.313 \\
FSI: Nucleon FSI ratio variation from nominal & 0 & 0.3 \\
\hline
\hline
\end{tabular}

\begin{tablenotes}
\footnotesize
\item[a] For the resonant $1\pi$ production process, the covariance matrix calculation in Ref.~\cite{odagawa_thesis} used central parameter values that were 4--5\% larger than the nominal values used to generate the MC prediction and the expected bin contents of the observables in this analysis. The nominal values were updated to follow the T2K oscillation analysis~\cite{t2k_epjc_2023} and are listed in Table~\ref{tab:neut_models}.
\end{tablenotes}
\end{threeparttable}
\end{table}

\subsubsection{Muon selection efficiency}
\label{sec:syst_mueff}

The dominant uncertainty arises from the background-corrected ST--LES track connection efficiencies shown in Fig.~\ref{fig:bm_les_ecc_eff}. The relatively large uncertainties at large muon angles are primarily due to the limited statistics of through-going muons available to constrain the efficiency in this region. In addition, a conservative 5\% normalization uncertainty was assigned to the overall muon selection efficiency to cover the observed data--MC differences in the BM--ST connection efficiency (2.2\% overall efficiency difference with additional angular dependence; Fig.~\ref{fig:bmst_eff}) and $O(0.1)\%$ efficiency reduction due to ST--LES track misconnection.

The efficiencies were varied within their assumed $1\sigma$ uncertainties, and the resulting changes in the predicted observable distributions were propagated to construct the corresponding systematic covariance matrices.

\subsubsection{Secondary interactions and detector response}

\label{sec:syst_det}

Systematic uncertainties arising from hadronic secondary interactions were evaluated by comparing different \textsc{Geant4} physics lists that implement distinct intranuclear cascade models for hadron–nucleus interactions in the $O(0.1\text{–}1)$~GeV energy range, namely \texttt{QGSP\_BIC} (Binary cascade model \cite{bic}) and \texttt{QGSP\_INCLXX} (Liège intranuclear cascade model \cite{incl1, incl2}).

Systematic uncertainties associated with the detector response were evaluated using both MC simulations and data. For the BM track reconstruction uncertainties, the simulation-based systematic uncertainty evaluation framework developed for the WAGASCI–BabyMIND analysis~\cite{wgbm_2025} was used. Notably, detector-related parameters, including the BM magnetic field strength, scintillator and readout properties, and the alignment of scintillator bars and detector components, were varied within their uncertainties as used in Ref.~\cite{wgbm_2025}. In addition, the basetrack angular resolution, detection efficiency, and the VPH probability density functions (PDFs) used for PID were varied according to the statistical uncertainties of their fits to data. Table~\ref{tab:det_syst} summarizes the considered detector response systematic uncertainties and their corresponding variations. The impact of each variation on the final observables was quantified through the resulting bin-by-bin covariance matrix.

\begin{table}[tb]
\centering
\caption{Detector-related systematic uncertainties considered in this analysis.}
\label{tab:det_syst}
\begin{tabularx}{\linewidth}{l X}
\hline\hline
Source & Variation / treatment \\
\hline

\multicolumn{2}{l}{\textbf{Physics modeling}} \\
Geant4 physics list &
QGSP\_BERT, BIC, and INCLXX \\
\hline

\multicolumn{2}{l}{\textbf{Detector geometry and response}} \\
BM--ST distance &
$\pm 1$ mm shift \\
ST SiPM dark noise hit probability &
0, 0.55, and 1.1\% per beam bunch \\
ST scintillator bar alignment &
2~mm offset between adjacent planes \\
ECC material thickness &
Steel/emulsion: $\pm0.3\%$; water: $\pm1.0\%$ \\
\hline

\multicolumn{2}{l}{\textbf{Reconstruction and calibration}} \\
BM track reconstruction \cite{wgbm_2025} &
Same variation as Ref.~\cite{wgbm_2025} \\
ST SiPM hit threshold &
1.5, 2.5, and 3.5 photoelectrons \\
Basetrack angular resolution &
Varied within fit uncertainty (statistical) \\
Basetrack detection efficiency &
Varied within statistical uncertainty \\
VPH PDF mean &
Varied within fit uncertainty (statistical) \\
VPH PDF width &
Varied within fit uncertainty (statistical) \\
Time-dependent VPH correction &
$\pm$10\% additional error in VPH PDF widths \\
MCS momentum scale correction \cite{odagawa_momentum} &
Enabled / disabled \\
\hline\hline
\end{tabularx}
\end{table}

\subsubsection{External beam-related backgrounds}
\label{sec:syst_bg}

The dominant background arises from neutrino interactions outside the ECC water fiducial volume that are misreconstructed as interactions within the fiducial volume. This contribution was estimated using an MC simulation that includes neutrino interactions both inside the ECC and in the surrounding detector materials and experimental hall. Some of these events are induced by neutral particles, such as neutrons and neutral kaons, produced in upstream interactions and subsequently interacting or decaying within the detector. In addition, charged particles originating outside the fiducial volume may undergo large-angle scattering in the ECC and be reconstructed as internal tracks, with the scattering point incorrectly identified as the interaction vertex. To account for potential mismodeling of these backgrounds, an 18\% normalization uncertainty was assigned based on the maximum discrepancy between the observed and predicted event yields across the event selection steps summarized in Table~\ref{tab:event_selection}.

The contribution from cosmic-ray muons coincident with the beam spill was neglected, as the expected number of such events in the analyzed data set was estimated to be significantly less than one after accounting for the detector acceptance, overburden, beam timing structure, and film tracking efficiency. The probability of an unrelated sand muon occurring within the same beam bunch was estimated to be approximately 0.7\% assuming Poisson statistics and was also neglected.

\subsection{Assigned fractional uncertainties to observable bins}

Figure \ref{fig:mult_syst} shows the fractional uncertainties assigned to each proton and pion multiplicity bin. For the 0-, 1-, and 2-track multiplicity bins, which contain the majority of selected events, the fractional uncertainties associated with beam flux, neutrino interaction, detector response, and muon selection efficiency modeling are relatively stable at approximately 5\%, 10--15\%, 10\%, and 10--15\%, respectively. The resulting total systematic uncertainty is typically 20--25\%, comparable to the statistical uncertainty in these bins.

\Fig{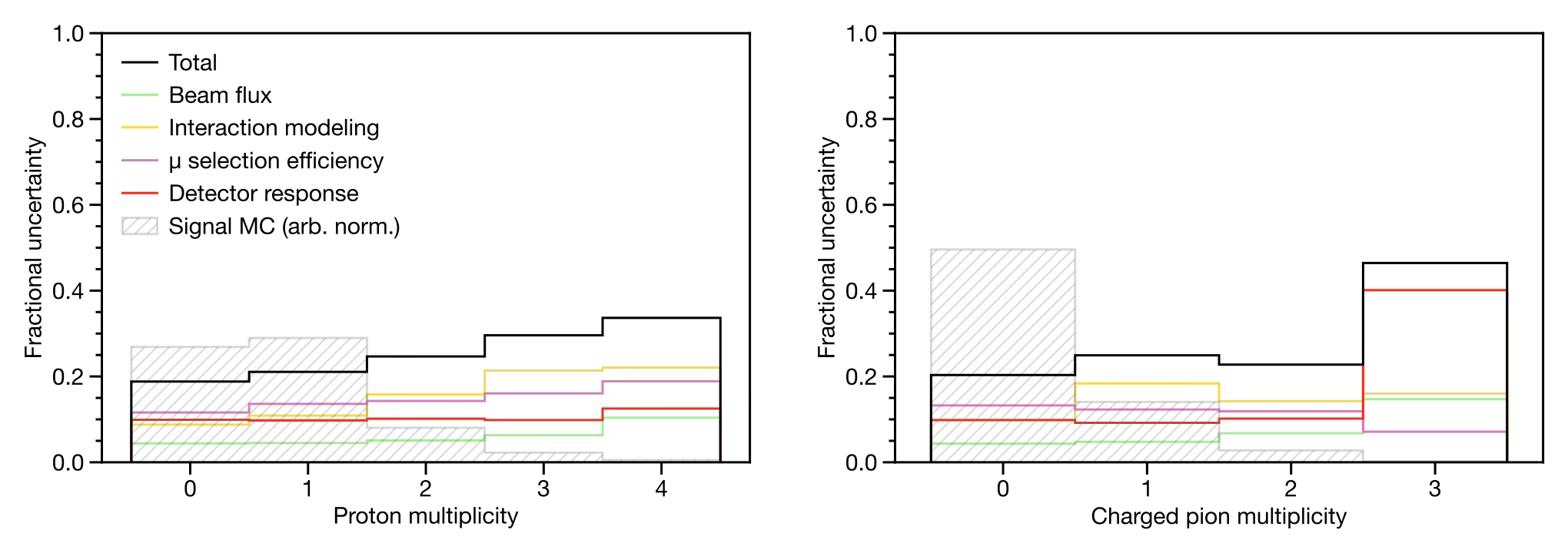}{\textwidth}{Bin-by-bin fractional uncertainties by source for the proton (left) and pion (right) multiplicity distributions. Dominant signal-related uncertainty sources are shown. The gray hatched histograms represent the expected signal ($\nu_\mu$ CC interactions on water) distributions in the MC simulation.}{fig:mult_syst}

Figure~\ref{fig:kin_syst} shows the fractional uncertainties assigned to each muon, proton, and pion kinematic bin. The muon angular distribution is primarily affected by the angle dependence of the muon connection efficiency, whereas detector response uncertainties dominate the reconstructed muon momentum. Neutrino interaction modeling is one of the dominant sources of uncertainty for the proton and pion angular distributions, particularly in the backward-scattering region.

\Fig{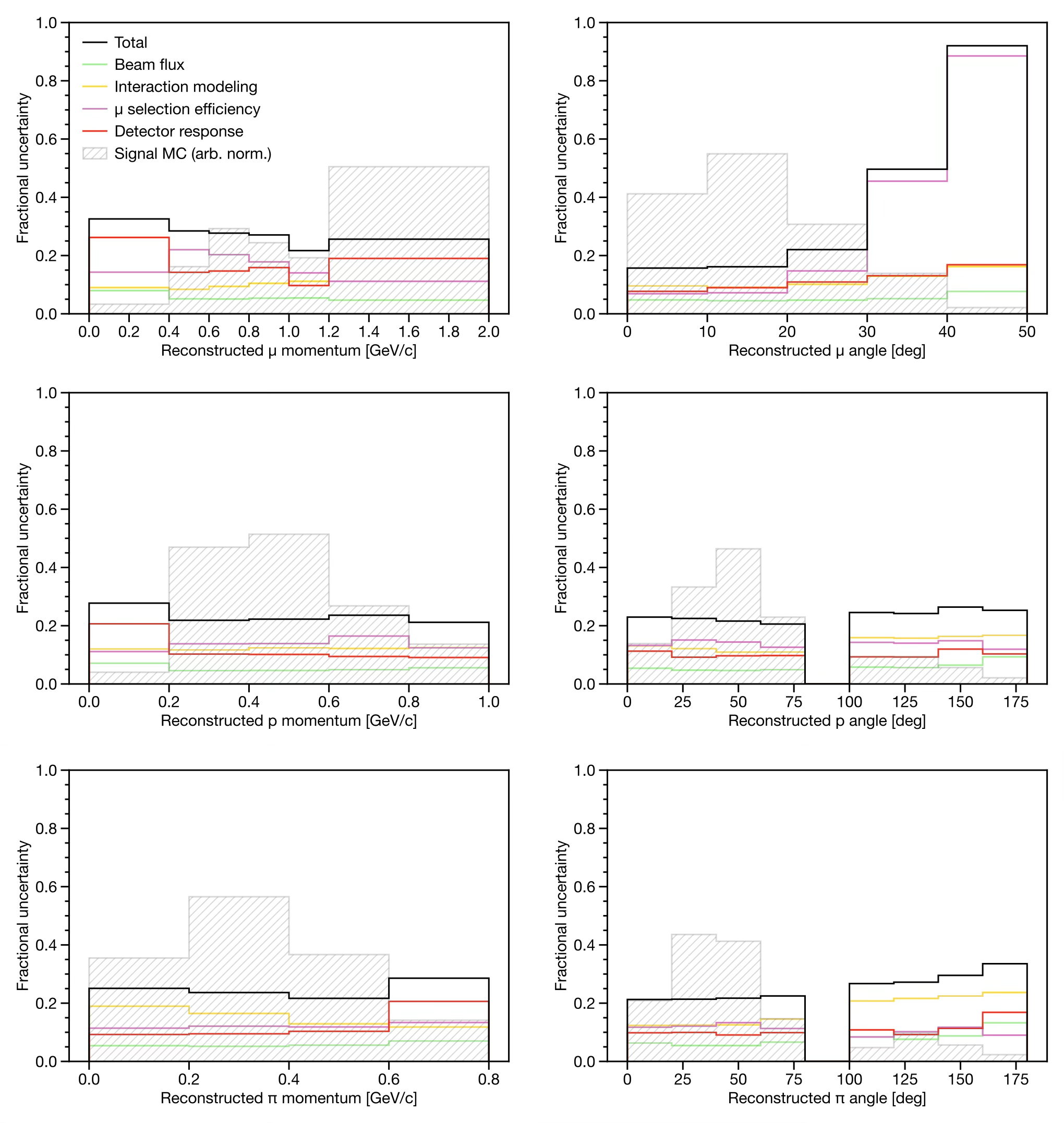}{\textwidth}{Bin-by-bin fractional uncertainties by source for muon momentum (top left), muon angle (top right), proton momentum (center left), proton angle (center right), charged pion momentum (bottom left), and charged pion angle (bottom right). Dominant signal-related uncertainty sources are shown. The gray hatched histograms represent the expected signal ($\nu_\mu$ CC interactions on water) distributions in the MC simulation.}{fig:kin_syst}

%% file: SubFiles/6_Results.tex
\section{Results}

In this section, reconstructed-level multiplicity and kinematic distributions for the selected flux-integrated inclusive CC $\nu_\mu$ event sample are compared with MC predictions. The predictions include the beam neutrino flux, neutrino interactions in the detector and surrounding materials, and the detector response simulation, as described in Sec.~\ref{sec:simulation}, and are normalized to the recorded beam POT. The selected sample is expected to have a signal purity of 80\% for $\nu_\mu$ CC interactions on water. The remaining 20\% consists of approximately 5\% wrong-sign (dominated by $\bar{\nu}_\mu$, with some $\nu_e$ and a negligible $\bar{\nu}_e$ contribution) and NC interactions on water, 8\% interactions in the tracking unit envelope, and 7\% external backgrounds from neutrino interactions occurring outside the ECC. No attempt was made to unfold the reconstructed distributions or extract interaction-level quantities. Track angles were measured with respect to the beam axis. The statistical uncertainty in each bin is typically around 20\%, comparable to the assigned total systematic uncertainty.

\subsection{Goodness-of-fit evaluation}
\label{sec:gof}

The goodness of fit between the reconstructed data and the MC prediction was evaluated using a profiled Poisson likelihood-ratio test statistic. The expected number of events in bin $i$ was parameterized as
\begin{equation}
\mu_i = m_i + \sum_k A_{ik} z_k ,
\end{equation}
where $m_i$ is the nominal MC prediction and the matrix $A_{ik}$ encodes the allowed systematic variations. It was obtained from the eigenvalue decomposition of the systematic covariance matrix $\mathbf{C}$, with $\mathbf{A}$ defined such that $\mathbf{C}=\mathbf{A}\mathbf{A}^T$. The index $k$ labels the independent eigenmodes of the covariance matrix. The parameters $z_k$ are nuisance parameters corresponding to independent covariance eigenmodes, and are constrained by standard Gaussian priors. The nuisance parameters were profiled by minimizing the penalized Poisson likelihood-ratio statistic
\begin{equation}
\chi^2_{\mathrm{P}}
=
2 \sum_i
\left[
\mu_i - d_i +
d_i \ln\!\left(\frac{d_i}{\mu_i}\right)
\right]
+
\sum_k z_k^2 ,
\label{eq:poisson_chi2}
\end{equation}
with respect to all $z_k$, where $d_i$ is the observed number of events in bin $i$. For bins with $d_i = 0$, the logarithmic term was taken to be zero. The minimum value of $\chi^2_{\mathrm{P}}$ obtained after profiling the nuisance parameters was used as the goodness-of-fit statistic.

The corresponding $p$-value was evaluated using toy MC pseudo-experiments. For each pseudo-experiment, nuisance parameters were first drawn from their standard Gaussian constraints, and the shifted expectation $\mu_i^{\rm toy}
= m_i + \sum_k A_{ik} z_k^{\rm toy}$
was constructed. Pseudo-data were then generated according to $d_i^{\rm toy} \sim \mathrm{Poisson}(\mu_i^{\rm toy})$. The same profiling procedure used for data was applied to each pseudo-experiment. The $p$-value was computed as the fraction of pseudo-experiments with a minimized $\chi^2_{\mathrm{P}}$ value greater than that observed in data.

The minimized goodness-of-fit statistic and corresponding $p$-value for each distribution are shown in Figs.~\ref{fig:total_mult}--\ref{fig:pi_kin}. Overall, the reconstructed distributions were found to be reasonably well described by the MC prediction. The resulting $p$-values indicate generally good agreement between data and simulation, although some distributions, such as the total charged particle multiplicity (Fig.~\ref{fig:total_mult}, $p=0.073$) and proton angular distribution (Fig.~\ref{fig:p_kin}, $p=0.011$), exhibit some tension with the prediction. No nuisance parameter was significantly constrained by the fit, with all nuisance vectors satisfying $\|\mathbf{z}\| < 1$. These conclusions were found to be stable under scaling variations of the assumed systematic uncertainties by $\pm 50\%$.

\subsection{Proton and charged pion multiplicities}

Figures~\ref{fig:total_mult} and \ref{fig:p_pi_mult} compare the total charged particle multiplicity, together with the proton and charged pion multiplicities for tracks with reconstructed $p\beta < 700$~MeV/$c$, between data and MC simulation. The total charged particle multiplicity exhibits a modest shape discrepancy, with the data generally favoring lower multiplicities than predicted by the MC. In contrast, the proton and charged pion multiplicity distributions are reasonably well reproduced by the simulation. For the total charged particle multiplicity, the profile likelihood fit favors a modest reduction of the higher-multiplicity events. This preference was primarily driven by the interaction model nuisance parameters, although no individual eigenmode was shifted by more than one standard deviation.

\Fig{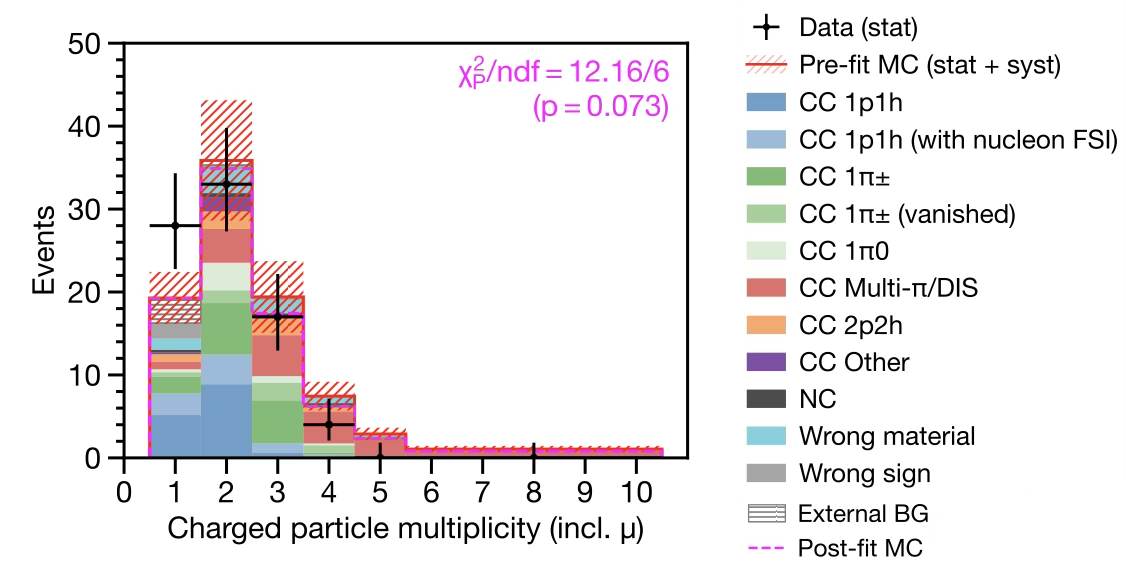}{0.7\textwidth}{Total charged particle multiplicity distribution, including unclassified tracks with $p\beta > 700$~MeV/$c$. Error bars on the data represent statistical uncertainties only, while the hatched bands on the MC prediction include both statistical and total systematic uncertainties described in Sec.~\ref{sec:syst}. The colored histograms show
the contributions from different simulated interaction channels for $\nu_\mu$
interactions on water. ``(with nucleon FSI)'' denotes events in which the outgoing nucleon undergoes final-state interactions within the nucleus, while ``(vanished)'' denotes events in which an outgoing charged pion is either absorbed or undergoes charge exchange to become a neutral pion. ``Wrong material'' refers to interactions occurring
in materials other than water (mostly the tracking unit envelope). ``Wrong
sign'' refers to water-target interactions induced by neutrino species other than
$\nu_\mu$ ($\bar{\nu}_\mu$, $\nu_e$, and $\bar{\nu}_e$). ``Post-fit MC'' denotes the prediction obtained after profiling the systematic nuisance parameters under Gaussian constraints. The quoted $p$-value is derived from the fraction of toy-MC pseudo-experiments with $\chi^2_\mathrm{P}$ values greater than that observed in data.
}{fig:total_mult}

\Fig{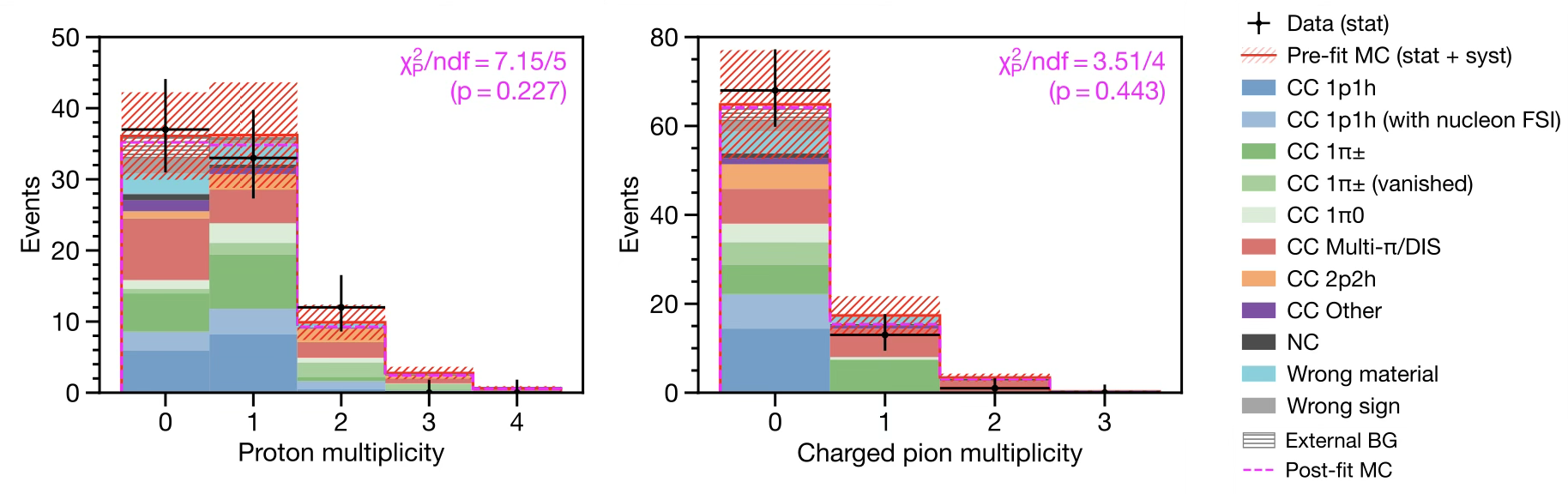}{\textwidth}{Proton multiplicities (left) and charged pion multiplicities (right).  Only tracks with reconstructed $p\beta < 700$~MeV/$c$ are shown, where PID is applied. The legend and uncertainty definitions are the same as those used in Fig.~\ref{fig:total_mult}.}{fig:p_pi_mult}

\subsection{Muon, proton, and charged pion kinematics}

Figures~\ref{fig:mu_kin}, \ref{fig:p_kin}, and \ref{fig:pi_kin} compare the reconstructed momentum and angular distributions of muons, protons, and charged pions between data and MC simulation. The MC-predicted purities of the selected muon, proton, and charged pion samples are 97\%, 99.7\%, and 94\%, respectively. A small excess of events is observed in the $30^\circ$--$40^\circ$ muon angular bin. The largest post-fit nuisance pull is associated with the muon connection efficiency, whose uncertainty increases at large angles because of the limited statistics of the large-angle sand muons used to determine the efficiency. This region will require further validation with higher-statistics data and improved modeling of the ST--LES connection efficiency, for example using a dedicated LES detector simulation instead of an efficiency lookup table derived from sand muon data.

The proton and charged pion kinematic distributions show good overall agreement with the MC prediction, except for the proton angle distribution where a deficit of forward-scattered protons is observed in the data. The profile likelihood fit favors a modest reduction in the predicted proton yield, with the dominant post-fit nuisance pulls arising from the interaction and detector response modeling.

Figure~\ref{fig:p_mom_vs_ang} compares the observed and predicted two-dimensional distribution of reconstructed momentum and angle for proton-like tracks. The compatibility of the observed and predicted distributions was evaluated using a KDE-based log-likelihood test calibrated with bootstrap pseudo-experiments, as shown in Fig.~\ref{fig:p_2d_corr}, yielding a $p$-value of 0.006 (see Sec.~\ref{sec:p_mom_vs_len} for the definition of the $p$-value). This low $p$-value was largely driven by a single data track with reconstructed momentum of 0.94~GeV/$c$ and angle of $108^\circ$, where removing this track increased the $p$-value to 0.283.

\Fig{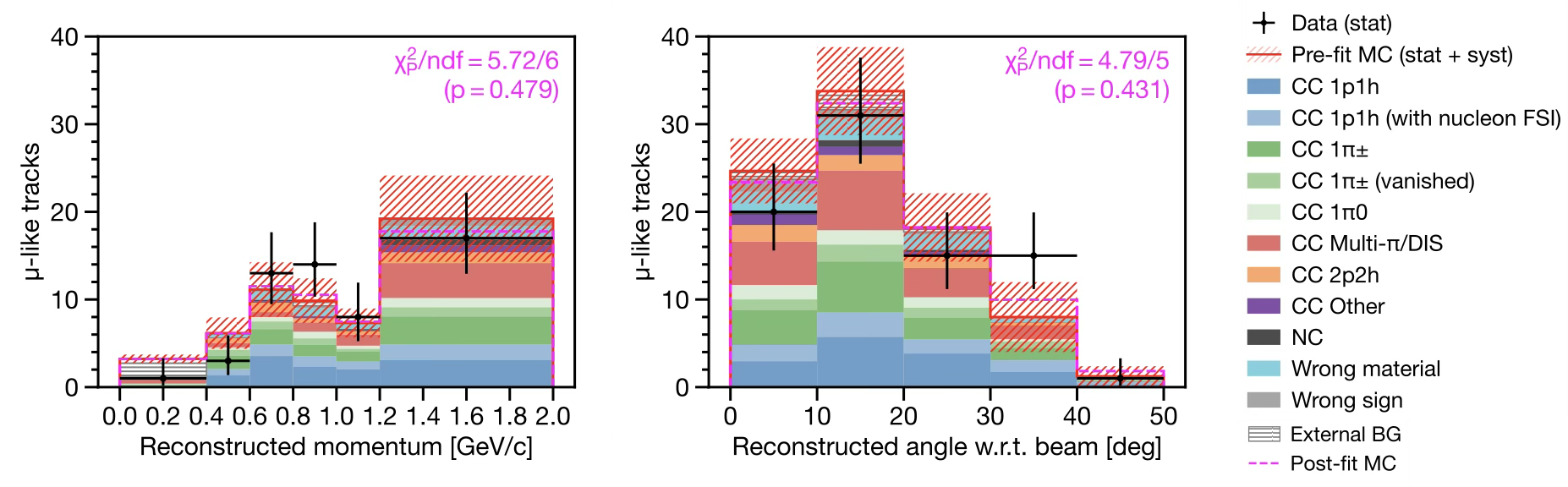}{\textwidth}{Reconstructed muon momentum in GeV/$c$ (left) and angle with respect to the beam (right). The legend and uncertainty definitions are the same as those used in Fig.~\ref{fig:total_mult}.}{fig:mu_kin}

\Fig{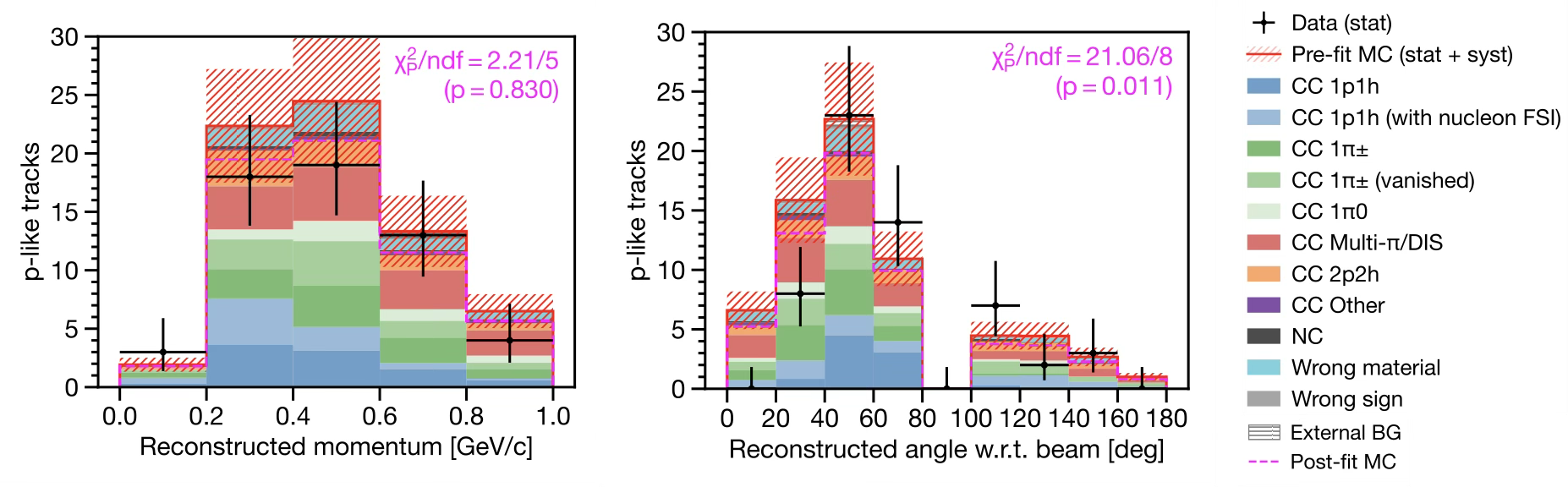}{\textwidth}{MCS-based reconstructed proton momentum (left) and angle with respect to the beam direction (right). Only tracks with reconstructed $p\beta < 700$~MeV/$c$ are shown, where PID is applied. The absence of events in the $80^\circ$--$100^\circ$ angular bin reflects the intrinsic acceptance of the ECC, which has limited sensitivity to tracks at large angles relative to the film surface. The legend and uncertainty definitions are the same as those used in Fig.~\ref{fig:total_mult}.}{fig:p_kin}

\Fig{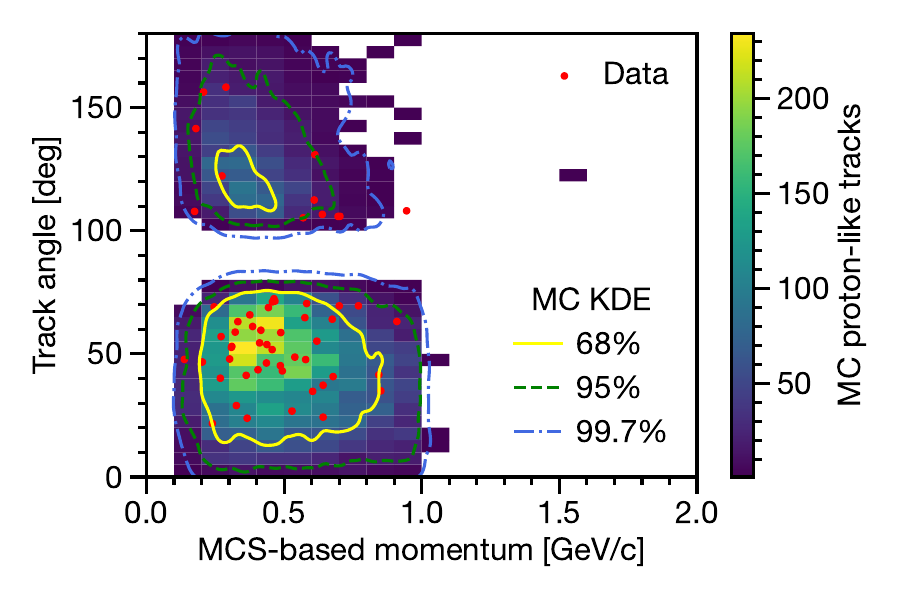}{0.6\textwidth}{Two-dimensional distribution of MCS-based reconstructed momentum versus track angle, for proton-like tracks shown in Fig.~\ref{fig:p_kin}. The colored contours show the 68\%, 95\%, and 99.7\% probability density regions obtained from a Gaussian KDE of the MC sample.}{fig:p_mom_vs_ang}

\Fig{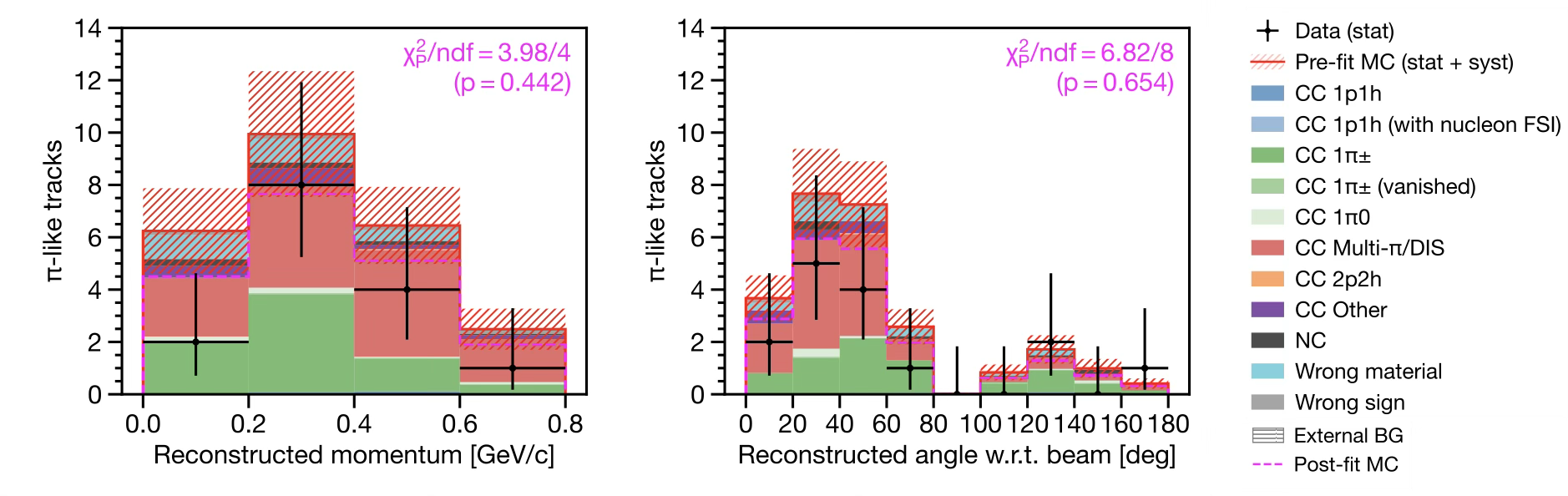}{\textwidth}{MCS-based reconstructed charged pion momentum (left) and angle with respect to the beam (right). Only tracks with reconstructed $p\beta < 700$~MeV/$c$ are shown, where PID is applied. The absence of events in the $80^\circ$--$100^\circ$ angular bin reflects the intrinsic acceptance of the ECC, which has limited sensitivity to tracks at large angles relative to the film surface. The legend and uncertainty definitions are the same as those used in Fig.~\ref{fig:total_mult}.}{fig:pi_kin}

%% file: SubFiles/7_Conclusions.tex
\section{Summary and prospects}

We have presented reconstructed-level final-state multiplicity and kinematic distributions for flux-integrated inclusive $\nu_\mu$ CC interactions on a water target using the NINJA emulsion-based detector system. The data correspond to an exposure of $4.63 \times 10^{20}$ POT and a fiducial mass equivalent to a 3.9~kg water target, representing approximately 5\% of the total neutrino interactions on water collected during the first NINJA physics run. The observed distributions were compared with reconstructed MC predictions based on the neutrino interaction models used in the T2K experiment. Most distributions were found to be in reasonable agreement within the statistical and estimated systematic uncertainties, although some tension was observed in the proton angular distribution with a deficit of forward-scattered protons. This analysis represents the first reconstructed-level characterization of proton kinematics down to 200~MeV/$c$ in $\nu_\mu$ interactions on water and establishes the reconstruction, detector simulation, and systematic uncertainty evaluation framework for analyses of the NINJA physics run data.

All three planned NINJA physics runs \cite{hayakawa_nufact, otani_nufact} have been completed, and the corresponding data will be analyzed using a similar framework. An increase in statistics by approximately one to two orders of magnitude is expected, enabling more precise studies of hadronic final states in neutrino interactions. The present analysis has also identified several areas where further improvements or verification are required, such as the muon connection efficiency at angles larger than $30^\circ$ and the observed deficit of forward-scattered protons. Future analyses of the complete NINJA data set will incorporate these improvements and validate the observed features using the substantially larger data sample, enabling more stringent tests of neutrino interaction models on water and reducing the associated interaction model uncertainties relevant to water-based long-baseline neutrino oscillation experiments.

%% file: SubFiles/Acknowledgment.tex
\section*{Acknowledgments}

We thank the T2K Collaboration for their strong support in conducting this experiment, as well as the J-PARC staff for their excellent accelerator performance. We also acknowledge the T2K WAGASCI/BabyMIND group for their stable operation, for providing experimental data, and for their cooperation in software development. We are grateful to the T2K neutrino beam group for delivering a high-quality beam and for their assistance with the beam MC simulation. We further thank the Instrument Development Center at Nagoya University for their valuable support in the development of the emulsion shifter.

This work was supported by the JST-SENTAN Program of the Japan Science and Technology Agency (JST), the Ministry of Education, Culture, Sports, Science and Technology of Japan (MEXT), and the Japan Society for the Promotion of Science (JSPS) through KAKENHI Grant Numbers JP17H02888, JP18H03701, JP18H05210, JP18H05535, JP18H05537, JP18H05541, JP20J15496, JP20J20304, and JP25K24555.

This work was also supported by the Croatian Science Foundation under the project numbers HRZZ-DOK-NPOO-2023-10-1262 and HRZZ-DOK-2025-02-2372; Swiss National Science Foundation and Croatian Science Foundation via grant MAPS IZ11Z0\_230193; Ministry of Science, Education and Youth of the Republic of Croatia via grant No. PK.1.1.10.0002 and via financial support for participating in the international NINJA project.


%

\vspace{0.2cm}
\noindent


\let\doi\relax